\documentclass[12pt]{article}

\usepackage{amsmath}
\usepackage{subfigure}
\usepackage{appendix,epic,eepic,amsmath,amsthm,amssymb,epsfig,cite,indentfirst,oldgerm}
\usepackage{multicol,boxedminipage}
\renewcommand{\theequation}{\arabic{equation}}
\newcommand{\EQ}{\begin{equation}}
\newcommand{\EN}{\end{equation}}

\newcommand{\bear}{\begin{eqnarray}}
\newcommand{\ear}{\end{eqnarray}}
\newcommand{\bt} { \begin{tabular} }
\newcommand{\et}{ \end{tabular} }
\newcommand{\bc} { \begin{center} }
\newcommand{\ec}{ \end{center} }

\newcommand{\btb} { \begin{table} }
\newcommand{\etb}{ \end{table} }

\begin{document}

\topmargin 0pt
\oddsidemargin 5mm
\newcommand{\NP}[1]{Nucl.\ Phys.\ {\bf #1}}
\newcommand{\PL}[1]{Phys.\ Lett.\ {\bf #1}}
\newcommand{\NC}[1]{Nuovo Cimento {\bf #1}}
\newcommand{\CMP}[1]{Comm.\ Math.\ Phys.\ {\bf #1}}
\newcommand{\PR}[1]{Phys.\ Rev.\ {\bf #1}}
\newcommand{\PRL}[1]{Phys.\ Rev.\ Lett.\ {\bf #1}}
\newcommand{\MPL}[1]{Mod.\ Phys.\ Lett.\ {\bf #1}}
\newcommand{\JETP}[1]{Sov.\ Phys.\ JETP {\bf #1}}
\newcommand{\TMP}[1]{Teor.\ Mat.\ Fiz.\ {\bf #1}}

\renewcommand{\thefootnote}{\fnsymbol{footnote}}

\newpage
\setcounter{page}{0}
\begin{titlepage}
\begin{flushright}

\end{flushright}
\vspace{0.5cm}
\begin{center}
{\large Integrable three-state vertex models with weights lying on genus five curves}\\
\vspace{1cm}
{\large M.J. Martins } \\
\vspace{0.15cm}
{\em Universidade Federal de S\~ao Carlos\\
Departamento de F\'{\i}sica \\
C.P. 676, 13565-905, S\~ao Carlos (SP), Brazil}\\
\vspace{0.35cm}
\end{center}
\vspace{0.5cm}

\begin{abstract}
We investigate the Yang-Baxter algebra 
for $\mathrm{U}(1)$ invariant three-state 
vertex models whose Boltzmann
weights configurations break explicitly  
the parity-time reversal symmetry. 
We uncover two families of regular Lax operators with  
nineteen non-null weights which ultimately 
sit on algebraic plane curves 
with genus five. We argue that these curves admit 
degree two morphisms onto elliptic curves and 
thus they are bielliptic. The associated $\mathrm{R}$-matrices 
are non-additive in the spectral parameters and it
has been checked that they satisfy the Yang-Baxter equation.
The respective 
integrable quantum spin-1 Hamiltonians are exhibited.
\end{abstract}

\vspace{.15cm} \centerline{}
\vspace{.1cm} \centerline{Keywords: Yang-Baxter Equation, Vertex Models, High Genus Curves}
\vspace{.15cm} \centerline{2013}

\end{titlepage}


\pagestyle{empty}

\newpage

\pagestyle{plain}
\pagenumbering{arabic}

\renewcommand{\thefootnote}{\arabic{footnote}}
\newtheorem{proposition}{Proposition}
\newtheorem{pr}{Proposition}
\newtheorem{remark}{Remark}
\newtheorem{re}{Remark}
\newtheorem{theorem}{Theorem}
\newtheorem{theo}{Theorem}

\def\ll{\left\lgroup}
\def\rr{\right\rgroup}

\newtheorem{Theorem}{Theorem}[section]
\newtheorem{Corollary}[Theorem]{Corollary}
\newtheorem{Proposition}[Theorem]{Proposition}
\newtheorem{Conjecture}[Theorem]{Conjecture}
\newtheorem{Lemma}[Theorem]{Lemma}
\newtheorem{Example}[Theorem]{Example}
\newtheorem{Note}[Theorem]{Note}
\newtheorem{Definition}[Theorem]{Definition}

\section{Introduction}

Over the past decades we have witnessed the importance played 
by vertex models in the development
of the theory of integrable systems in two spatial dimensions \cite{BAX}. Let us
recall some of the
basic notions about this classical lattice model of statistical mechanics. For simplicity
consider a square lattice of size $\mathrm{N} \times \mathrm{N}$ with periodic boundary conditions
on both horizontal and vertical directions. The statistical configurations are 
specified by assigning to each set of four 
edges meeting at a given lattice site the spins variables $\alpha,\beta,\gamma$ and $\delta$. Here we assume that
these state variables take values on a finite subset $\mathrm{V}$ of the integers, that is
$\mathrm{V}=\{1,\cdots,q\}$. To a given vertex of the lattice we assign a Boltzmann weight
$\mathrm{W}_{\alpha,\beta}^{\gamma,\delta}$ representing the energy of 
the corresponding spins configurations.  
In figure \ref{weight} we illustrated our
notation for the weight indices.
\setlength{\unitlength}{900sp}
\begin{figure}[ht]
\begin{center}
\begin{picture}(6000,10000)(0,0)
\put(-50,4000){\makebox(0,0){$\mathrm{W}_{\alpha,\beta}^{\gamma,\delta}~=$}}
\put(2500,4000){\line(1,0){3000}}
\put(5500,4000){\line(1,0){3000}}
\put(5500,1000){\line(0,1){3000}}
\put(5500,4000){\line(0,1){3000}}
\put(2800,4350){\makebox(0,0){$\alpha$}}
\put(8200,4350){\makebox(0,0){$\gamma$}}
\put(5850,6900){\makebox(0,0){$\delta$}}
\put(5150,1300){\makebox(0,0){$\beta$}}
\end{picture} \par
\end{center}
\caption{The Boltzmann weight of the vertex model on a square lattice.}\label{weight}
\end{figure}

One feature of the vertex models is that its basic 
properties can be formulated
with the help of a beneath tensor structure 
formally  represented by the
product $\mathrm{V}_0 \otimes 
\mathrm{V}_k$ \cite{TATAFA}.  The horizontal degrees of freedom are encoded in the term   
$\mathrm{V}_0$  which is often called auxiliary space.
The second term $V_k$ 
stands for the vertical degrees of freedom associated to each $k$-th site of a
one dimensional lattice of size 
$\mathrm{N}$ playing the role of the quantum space of a $q$-state spin chain.
It turns out that the Boltzmann weights can be organized in terms of a local matrix 
acting on $\mathrm{V}_0  
\otimes \mathrm{V}_k$  
denominated Lax operator, 
\EQ
{\mathrm{L}}_{k}({\mathbf{w}})= \sum_{\alpha,\beta,\gamma,\delta=1}^{q} 
\mathrm{W}_{\alpha,\beta}^{\gamma,\delta}~~e_{\alpha \gamma}^{(0)} \otimes e_{\beta \delta}^{(k)},~~\mathrm{for}~~k=1,\cdots,\mathrm{N},
\EN
where 
$e_{\alpha \beta}^{(j)}$ denotes $q \times q$ Weyl matrices acting
on the space $\mathrm{V}_{j}$ with $j=0,\cdots,\mathrm{N}$. From now on we shall 
refer to the set of nonzero weights
$\mathrm{W}_{\alpha,\beta}^{\gamma,\delta}$ using the symbol $``{\mathbf{w}}"$. 

The respective row-to-row transfer matrix $\mathrm{T}({\bf w})$ can then be written
as the trace over the auxiliary space of an ordered product of Lax operators, namely
\EQ
\mathrm{T}(\mathbf{w})=\mathrm{Tr}_{\mathrm{V}_0} 
\left[ {\mathrm{L}}_{\mathrm{N}}({\mathbf{w}}) 
{\mathrm{L}}_{\mathrm{N-1}}({\mathbf{w}}) 
\cdots
{\mathrm{L}}_{2}({\mathbf{w}}) 
{\mathrm{L}}_{1}({\mathbf{w}}) \right].
\label{TRA}
\EN

A relevant family of vertex models are those whose Lax operators are invariant by a single
$\mathrm{U}(1)$ symmetry. This invariance implies that many of the weights
are zero depending on whether or not the respective indices 
satisfy the so-called ice condition, 
\begin{eqnarray}
\label{ICE}
&& \bullet \mathrm{W}_{\alpha,\beta}^{\gamma,\delta}=0,~~\mathrm{for}~~\alpha+\beta \neq \gamma+\delta \nonumber \\
&& \bullet \mathrm{W}_{\alpha,\beta}^{\gamma,\delta} \neq 0,~~\mathrm{for}~~\alpha+\beta = \gamma+\delta. 
\end{eqnarray}

Up to the present the known realizations of
vertex models satisfying the above rule have the corresponding weights parameterized
in terms of trigonometric functions.
The typical examples are the vertex models associated to the solutions of the Yang-Baxter
equation based on the quantum $\mathrm{U}[\mathrm{SU}(2)]_{\bar{q}}$ algebra either for generic values of
the deformation parameter \cite{ZAFASO,BARE} or when it 
takes values on the roots of unity \cite{DEG,GOM}. 
The current results in
the literature suggest that to obtain  
integrable vertex models with weights not uniformized by rational functions one has to
consider statistical configurations
that violate the $\mathrm{U}(1)$ symmetry. For example, these are the cases 
of certain generalizations of the eight-vertex model \cite{BELA} 
and the celebrated chiral Potts model 
\cite{CH,CH1} having both an underlying $\mathbb{Z}_q$ symmetry.

On the other hand, it has been shown 
that the transfer matrix 
of $\mathrm{U}(1)$ invariant
vertex models can be diagonalizable 
by the algebraic Bethe ansatz for arbitrary 
Lax operators without reference to a given specific 
parameterization of the weights \cite{MEMA}. In this work it was not needed to make
any assumption on the dependence of the spectral 
parameters entering the corresponding $\mathrm{R}$-matrix to build 
up the transfer matrix eigenvectors. In any way this algebraic  approach  
forbids the existence of vertex models
having both the $\mathrm{U}(1)$ invariance and Boltzmann
weights sitting on algebraic varieties which can not be 
rationally uniformized. In fact, the generality  
of a number of weights identities derived in \cite{MEMA} 
make it hard to believe
that they are only realized in terms of trigonometric functions. 

Of course, for such class of models, irrationality of weights
can only emerge when the number of states $q$ is larger than two. 
This is because the model with $q=2$ corresponds 
to the asymmetric six-vertex model whose weights are 
known to be rationally parameterized, see for 
example \cite{BAX1}. For $q >2$, however, the structure of the
functional relations  derived from the Yang-Baxter algebra
is very different from that satisfied by weights of an arbitrary six-vertex model.
Would they be so stringent to
always drive us to exactly solvable vertex models with trigonometric weights?. Here we investigate 
this question in the most simple case
where non-rational weights can not be rule out: the three-state $\mathrm{U}(1)$ vertex model.
Another relevant motivation to
study these kind of systems comes from the existence of 
concrete exactly solvable spin-$1$ quantum chains discovered
by Alcaraz and Bariev within the coordinate Bethe ansatz method \cite{ALBA}.
The fact that their
Hamiltonian for general couplings can not be derived in terms of 
an additive $\mathrm{R}$-matrix suggests that non rational 
three-state vertex models should indeed exist.

Our study of Yang-Baxter algebra for $\mathrm{U}(1)$ three-state vertex
models will lead us to develop a strategy to deal with a problem involving 
a large number of functional relations constraining the Boltzmann weights.
The basic guidelines of our approach is somehow general 
and can in principle be used to study more complicated vertex models.
It turns out that we are able to uncover two families 
of integrable $\mathrm{U}(1)$ three-state vertex models 
with weights lying on non-rational manifolds. In fact, we shall argue that their 
Boltzmann weights are ultimately constrained 
by algebraic plane curves of genus five. We remark that the respective quantum spin-$1$ chains 
extend in a substantial way the previous integrable Hamiltonian 
found in reference \cite{ALBA}.

We have organized this work as follows. In next section we describe
some basic properties of the polynomial relations coming from the
Yang-Baxter algebra. We have endeavored to make it self contained when
useful mathematical notions of algebraic geometry are used.
In section \ref{vertexmodel} we describe the main structure of the
$\mathrm{U}(1)$ invariant three-state vertex model to be
studied in this paper. The solution of the
respective functional relations is
detailed in Section \ref{funrel} and we have been able to
uncover two families of integrable three-state vertex models.
In Section \ref{geoman} we investigate the
geometrical properties underlying the integrability of these
vertex models and for one of the families this forces us to 
analyze the problem of the intersection of two projective surfaces. 
We found that the
properties of the underlying algebraic manifolds are related to that
of genus five bielliptic curves. We have summarized the
main results for the Lax operators and the respective
$\mathrm{R}$-matrix in Section \ref{LAXRH}. We compute
the expressions of the corresponding exactly solvable spin-$1$ chains and
show that they contain as particular case the spin-$1$ Hamiltonian found 
previously by Alcaraz and Bariev \cite{ALBA}. Our conclusions are
presented in Section \ref{conclu} and in four Appendices we describe
a number of technical details complementing the discussions of 
the main text.

\section{Integrability Conditions}

In general, a lattice model of statistical mechanics in two-dimensions 
is considered integrable when the
corresponding transfer matrix can be embedded 
into a family of pairwise commuting operators \cite{BAX},
\EQ
\left [ \mathrm{T}({\mathbf w}^{'}), \mathrm{T}({\mathbf w}^{''}) \right ]=0, 
\label{COM}
\EN
where ${\mathbf w}^{'}$ and ${\mathbf w}^{''}$ 
represent two different sets of weights. 

A sufficient condition for commuting transfer matrices 
was originally introduced by Baxter
in his analysis of the eight-vertex model \cite{BAX2}. This condition requires 
the existence of an
invertible $\mathrm{R}$-matrix which together with the Lax operators should satisfy
the Yang-Baxter algebra,
\EQ
\mathrm{R}({\mathbf{w}}^{'},{\mathbf{w}}^{''}) [\mathrm{L}_k({\mathbf{w}}^{'}) \otimes \mathrm{I}_q]
[\mathrm{I}_q \otimes \mathrm{L}_k({\mathbf{w}}^{''})] =
[\mathrm{I}_q \otimes \mathrm{L}_k({\mathbf{w}}^{''})] [\mathrm{L}_k({\mathbf{w}}^{'}) \otimes \mathrm{I}_q]
\mathrm{R}({\mathbf{w}}^{'},{\mathbf{w}}^{''}),
\label{YB}
\EN
where $\mathrm{I}_q$ denotes the $q \times q $ identity and 
$\mathrm{R}({\mathbf{w}}^{'},{\mathbf{w}}^{''})$ is a $q^2 \times q^2$ matrix acting on the tensor
product $\mathrm{V}_0 \otimes \mathrm{V}_0$.

The commutation relation of two  
distinct transfer matrices (\ref{COM}) is obviously not affected 
when their weights  are multiplied by 
two independent nonzero scalar factors. 
This means that the functional equations coming from Yang-Baxter algebra
are expected to be homogeneous separately in each of the sets 
${\mathbf{w}}^{'}$ and ${\mathbf{w}}^{''}$ of weights. 
To be more precise
writing $\mathrm{F}_j({\mathbf{w}}^{'},{\mathbf{w}}^{''})$ to denote
a given polynomial derived from Eq.(\ref{YB})
we then, in general, can state that,
\EQ
\mathrm{F}_j(\lambda_1 {\mathbf{w}}^{'},\lambda_2{\mathbf{w}}^{''})  
=\lambda_1^{D_1} \lambda_2^{D_2} \mathrm{F}_j({\mathbf{w}}^{'},{\mathbf{w}}^{''}),~\forall~\lambda_1,\lambda_2 \neq 0,  
\EN
where $D_1$ and $D_2$ define the bidegree of the bihomogeneous 
polynomial $\mathrm{F}_j({\mathbf{w}}^{'},{\mathbf{w}}^{''})$.  

In this paper we will also assume that the 
$\mathrm{R}$-matrix satisfies the standard
unitarity condition,
\EQ
\mathrm{R}({\mathbf{w}}^{'},{\mathbf{w}}^{''}) 
\mathcal{P} \mathrm{R}({\mathbf{w}}^{''},{\mathbf{w}}^{'})
\mathcal{P}=  
\rho({\mathbf{w}}^{'},{\mathbf{w}}^{''})  \mathrm{I_q} \otimes \mathrm{I_q}
\label{UNI}
\EN
where $\mathcal{P}$ denotes the permutator operator 
acting on a $q^2$-dimensional space and 
$\rho({\mathbf{w}}^{'},{\mathbf{w}}^{''})$ represents an overall normalization.

The above assumption is motived by the fact that 
unitarity property (\ref{UNI}) assures us from the
very beginning that the $\mathrm{R}$-matrix has an inverse. Recall that unitarity
has also been relevant in providing us a number of identities
that were essential for the algebraic diagonalization of the 
transfer matrix of the $U(1)$ invariant vertex
models \cite{MEMA}.
In addition, we shall show that the unitarity property of the $\mathrm{R}$-matrix imposes 
an important restriction on the
structure of the polynomials 
$\mathrm{F}_j({\mathbf{w}}^{'},{\mathbf{w}}^{''})$. 
In order to see
that we multiply the left and right sides of Eq.(\ref{YB}) by the 
inverse of the $\mathrm{R}$-matrix
and with the help of Eq.(\ref{UNI}) we obtain,  
\EQ
[\mathrm{L}_k({\mathbf{w}}^{'}) \otimes \mathrm{I}_q] [\mathrm{I}_q \otimes \mathrm{L}_k({\mathbf{w}}^{''})]
\mathcal{P} \mathrm{R}({\mathbf{w}}^{''},{\mathbf{w}}^{'}) \mathcal{P}=
\mathcal{P} \mathrm{R}({\mathbf{w}}^{''},{\mathbf{w}}^{'}) \mathcal{P}
[\mathrm{I}_q \otimes \mathrm{L}_k({\mathbf{w}}^{''})][\mathrm{L}_k({\mathbf{w}}^{'}) \otimes \mathrm{I}_q].
\label{YB1}
\EN

We now apply the permutator on both sides of Eq.(\ref{YB1}) as well
as we insert the identity $\mathcal{P}^2=\mathrm{I}_q \otimes \mathrm{I}_q$ in the middle of the brackets
to permute the Lax operators.
As a result we can derive the following relation,
\EQ
[\mathrm{I_q} \otimes \mathrm{L}_k({\mathbf{w}}^{'})][ \mathrm{L}_k({\mathbf{w}}^{''}) \otimes \mathrm{I}_q]
\mathrm{R}({\mathbf{w}}^{''},{\mathbf{w}}^{'})=
\mathrm{R}({\mathbf{w}}^{''},{\mathbf{w}}^{'})
[\mathrm{L}_k({\mathbf{w}}^{''}) \otimes \mathrm{I}_q][\mathrm{I}_q \otimes \mathrm{L}_k({\mathbf{w}}^{'})].
\label{YB2}
\EN

Inspecting Eqs.(\ref{YB},\ref{YB2}) we see that their left and right sides are related
once we interchange 
the weights, that is ${\mathbf{w}}^{'} \leftrightarrow {\mathbf{w}}^{''}$. This means 
that the polynomial equations coming
from the Yang-Baxter algebra are expected to be anti-symmetrical 
upon the exchange of weights label, that is, 
\EQ
\mathrm{F}_j({\mathbf{w}}^{'},{\mathbf{w}}^{''})+
\mathrm{F}_j({\mathbf{w}}^{''},{\mathbf{w}}^{'})=0.
\label{ANTI}
\EN

We stress that such simple consequence of the unitarity of the $\mathrm{R}$-matrix is
going to play an important role to help us disentangle involved high degree functional
relations on the Boltzmann weights. It is however fortunate that in 
many instances of our analysis this fact will come out naturally since
we will be able to write the polynomials
in the following
particular anti-symmetrical form,
\EQ
\mathrm{F}_j({\mathbf{w}}^{'},{\mathbf{w}}^{''})=
\mathrm{H}_j({\mathbf{w}}^{'})
\mathrm{G}_j({\mathbf{w}}^{''})-
\mathrm{H}_j({\mathbf{w}}^{''})
\mathrm{G}_j({\mathbf{w}}^{'}),
\label{fact}
\EN
where 
$\mathrm{H}_j(\mathbf{w})$ and 
$\mathrm{G}_j(\mathbf{w})$ are irreducible
homogeneous polynomials with the same degree $D$.  
We then say that 
$\mathrm{F}_j({\mathbf{w}}^{'},{\mathbf{w}}^{''})$ 
is an irreducible bihomogeneous 
polynomial with bidegree $(D,D)$.

We see that polynomials of the form (\ref{fact}) 
vanish trivially when we consider the limit  
${\mathbf{w}}^{'} \rightarrow {\mathbf{w}}^{''}$ 
which is a desirable property since certainly 
the transfer matrix commutes with itself.  
We next note that such bihomogeneous polynomials 
always admit an special  
solution in which the distinct group of weights
${\mathbf{w}}^{'}$ and ${\mathbf{w}}^{''}$ 
are decoupled from each other. This solution bears some 
resemblance with the method of separation of variables used to solve the
classical dynamics by the Hamilton-Jacobi theory and 
partial differential equations of mathematical physics. It can be
written as follows,
\EQ
\frac{\mathrm{H}_j({\mathbf{w}}^{'})}{
\mathrm{G}_j({\mathbf{w}}^{'})}=
\frac{\mathrm{H}_j({\mathbf{w}}^{''})}{
\mathrm{G}_j({\mathbf{w}}^{''})}= \Lambda_j,
\label{factD}
\EN
where the parameter $\Lambda_j$ is considered 
a free constant. 

It turns out that such particular 
solution to Eq.(\ref{fact}) has a very clear
meaning in the realm of algebraic geometry. This discussion permits us
to introduce the appropriate mathematical terminology making 
presentation self contained.
We start by recalling that 
the zero locus of the bihomogeneous polynomial 
$\mathrm{F}_j({\mathbf{w}}^{'},{\mathbf{w}}^{''})$ is known 
to produce a well defined algebraic variety $\mathrm{X}_j$ \cite{SHAFA}.
In order to define this mathematical object  
let us denote the set of weights by the elements
$\omega_0,\hdots,\omega_m$ representing the coordinates of 
a projective space $\mathbb{CP}^m$ over the complex field.
The algebraic variety $\mathrm{X}_j$ is a closed subset 
of the product of such two projective spaces 
which formally can be represented as,
\begin{eqnarray}
\label{projset1}
\mathrm{X}_j&=&\{[\omega_{0}^{'}:\hdots:\omega_m^{'}]
\times [\omega_{0}^{''}:\hdots:\omega_m^{''}] \in
\mathbb{CP}^m\times \mathbb{CP}^m | \mathrm{H}_j(\omega_0^{'},\hdots,\omega_m^{'}) 
\mathrm{G}_j(\omega_0^{''},\hdots,\omega_m^{''}) \nonumber \\
&-&\mathrm{H}_j(\omega_0^{''},\hdots,\omega_m^{''}) 
\mathrm{G}_j(\omega_0^{'},\hdots,\omega_m^{'})=0
\}, 
\end{eqnarray}
where $[\omega_0:\hdots:\omega_m]$ denotes a point in 
the projective space $\mathbb{CP}^m$ by which we mean the line spanned
by the vector $(\omega_0,\hdots,\omega_m) \in \mathbb{C}^{m+1}$ 
where the origin is omitted. 

By the same token, the polynomials originated from the 
special solution (\ref{factD}) can also be used to define an underlying
subvariety $\mathrm{Y}_j \subset \mathrm{X}_j$. This subvariety is
in fact described by the product of two identical algebraic sets since the corresponding
polynomials do not mix distinct weights labels. 
We then are able to write $\mathrm{Y}_j = \mathrm{\overline{Y}}(\Lambda_j) 
\times \mathrm{\overline{Y}}(\Lambda_j)$ where
the component $\mathrm{\overline{Y}}(\Lambda_j)$ is defined by, 
\EQ
\label{projset2}
\mathrm{\overline{Y}}(\Lambda_j)=\{[\omega_{0}:\hdots:\omega_m]
\in \mathbb{CP}^m | \mathrm{H}_j(\omega_0,\hdots,\omega_m)-\Lambda_j \mathrm{G}_j(\omega_0,\hdots,\omega_m)=0 
\} 
\EN

We now can show that the particular solution (\ref{factD}) gives rise to a divisor
on the original variety $\mathrm{X}_j$. To this end we recall that one basic invariant of
any variety is its dimension which here can be determined using the standard result
that an irreducible hypersurface $\mathrm{S}(w_0,\hdots,w_m) \in \mathbb{CP}^m$ has 
dimension $\mathrm{dim S}=m-1$ \cite{SHAFA}. From this result it follows that
the variety $X_j$ has dimension $\mathrm{dimX}_j=2m-1$  while 
the dimension of the subvariety $\mathrm{Y}_j$ is 
$\mathrm{dimY}_j=2m-2$ since they are 
generated by irreducible polynomials. From the fact that $\mathrm{dim X}_j-\mathrm{dim Y}_j=1$  
and observing that the intersection multiplicity of
$\mathrm{Y}_j$ at $\mathrm{X}_j$ is also $1$ we then
conclude that $\mathrm{Y}_j$ is in fact a prime divisor 
element on $\mathrm{X}_j$. 
This means that by
varying the parameter $\Lambda_j$ we are able to foliate the variety
$\mathrm{X}_j$ through submanifolds of codimension $1$ whose fibers
are determined by the variety $\mathrm{\overline{Y}}(\Lambda_j)$.

We shall see that the prime divisors associated to 
polynomials with the structure (\ref{fact}) are precisely the
fundamental building blocks of a vertex model with commuting
transfer matrix. 
It is the intersection
of a collection of such divisors that ultimately is going to dictate 
the algebraic variety in which the Boltzmann weights are lying on.
This procedure assures us the existence of two independent transfer matrices
that are sited on the same algebraic manifold and thus of a single
family of Lax operators.

\section{Three-state Vertex Model }
\label{vertexmodel}

We now turn our attention to the presentation of 
the specific $\mathrm{U}(1)$ three-state vertex which
we intend to investigated in this paper.  From the ice-rule (\ref{ICE}) 
it follows that this type of model can have 
at most nineteen different Boltzmann weights. This space of parameters can be
reduced once we consider typical symmetries 
of the weights when they are viewed as $(1+1)$-dimensional scattering
amplitudes \cite{ZAMO}. These invariances are defined as follows,
\begin{eqnarray}
&& \bullet \mathrm{Parity~Reversal}: \mathrm{W}_{\alpha,\beta}^{\gamma,\delta}= \mathrm{W}_{\beta,\alpha}^{\delta,\gamma}, \nonumber \\
\label{SIM}
&& \bullet \mathrm{Time~Reversal}: \mathrm{W}_{\alpha,\beta}^{\gamma,\delta}= \mathrm{W}_{\gamma,\delta}^{\alpha,\beta}, \\
&& \bullet \mathrm{Charge~Conjugation}: \mathrm{W}_{\alpha,\beta}^{\gamma,\delta}= \mathrm{W}_{q+1-\alpha,q+1-\beta}^{q+1-\gamma,q+1-\delta}. \nonumber 
\end{eqnarray}

According to the recent work \cite{ROMA}  
nineteen vertex models invariant by the combined action
of parity and time reversal symmetries have always rational weights.
This means that we have to consider vertex models 
whose statistical configurations
do not preserve the $\mathrm{PT}$ transformation.
From Eqs.(\ref{SIM}) we see that this is 
achieved when the respective 
weights fulfill one of the following inequalities,
\EQ
\mathrm{W}_{12}^{12} \neq \mathrm{W}_{21}^{21},~~
\mathrm{W}_{13}^{13} \neq \mathrm{W}_{31}^{31},~~
\mathrm{W}_{23}^{23} \neq \mathrm{W}_{32}^{32},~~
\mathrm{W}_{13}^{22} \neq \mathrm{W}_{22}^{31},~~
\mathrm{W}_{22}^{13} \neq \mathrm{W}_{31}^{22}.
\label{PTS}
\EN

One way to assure the  
breaking of the $\mathrm{PT}$ 
symmetry is by means the diagonal weights 
$\mathrm{W}_{\alpha,\beta}^{\alpha,\beta}$ since the off-diagonal ones can 
in principle be modified
with the help of gauge transformations. In this case, broken
$\mathrm{PT}$ invariance is not completely incompatible 
with the preservation of charge conjugation  which
in turn permits us to work with a smaller number of
distinct weights. Considering that charge symmetry is
preserved at least by the diagonal weights our
starting ansatz for
the Lax operator is,
\EQ
\mathrm{L_k}({\mathbf{w}})=\left[
\begin{array}{c|c|c}
a~e_{11}^{(k)} +b~e_{22}^{(k)} +f~e_{33}^{(k)}  &
c~e_{21}^{(k)} +d~e_{32}^{(k)} & h~e_{31}^{(k)} \\ \hline
c~e_{12}^{(k)} +\bar{d}~e_{23}^{(k)}  
& \bar{b}~e_{11}^{(k)} +g~e_{22}^{(k)} +\bar{b}~e_{33}^{(k)} 
& d~e_{21}^{(k)} +c~e_{32}^{(k)} \\ \hline
\bar{h}~e_{13}^{(k)} &
\bar{d}~e_{12}^{(k)} +c~e_{23}^{(k)} & 
f~e_{11}^{(k)} +b~e_{22}^{(k)} +a~e_{33}^{(k)} \\
\end{array}
\right],
\label{LAX}
\EN
where $a$, $b$, $\bar{b}$, $c$, $d$, $\bar{d}$, $f$, $g$, $h$ and $\bar{h}$
denote ten distinct weights
of the set ${\mathbf{w}}$.
We see that the $\mathrm{PT}$ invariance is only broken 
by way of the diagonal weights
$b$ and $\bar{b}$. 

We now consider an arbitrary 
$\mathrm{R}$-matrix  and substitute it together with the
above ansatz for the Lax operators in the Yang-Baxter algebra (\ref{YB}).
It is not difficult to see that the underlying 
$\mathrm{U}(1)$ invariance of the Lax operators imposes 
us severe constraints 
on the $\mathrm{R}$-matrix. Under the mild assumption 
that some of the weights of the Lax operators  are not trivially related  we find that 
$\mathrm{R}$-matrix 
elements have also to satisfy the ice-rule (\ref{ICE}). This motivates us to choose
the $\mathrm{R}$-matrix with the same structure of the  Lax operators, namely
\EQ
\mathrm{R}({\mathbf{w}}^{'},{\mathbf{w}}^{''})=\left[
\begin{array}{ccc|ccc|ccc}
 \mathbf{a} & 0 & 0 & 0 & 0 & 0 & 0 & 0 & 0 \\
 0 & \mathbf{b} & 0 & \mathbf{c} & 0 & 0 & 0 & 0 & 0 \\
 0 & 0 & \mathbf{f} & 0 & \mathbf{d} & 0 & \mathbf{h} & 0 & 0 \\ \hline
 0 & \mathbf{c} & 0 & \mathbf{\bar{b}} & 0 & 0 & 0 & 0 & 0 \\
 0 & 0 & \mathbf{\bar{d}} & 0 & \mathbf{g} & 0 & \mathbf{d} & 0 & 0 \\
 0 & 0 & 0 & 0 & 0 & \mathbf{\bar{b}} & 0 & \mathbf{c} & 0 \\\hline
 0 & 0 & \mathbf{\bar{h}} & 0 & \mathbf{\bar{d}} & 0 & \mathbf{f} & 0 & 0 \\
 0 & 0 & 0 & 0 & 0 & \mathbf{c} & 0 & \mathbf{b} & 0 \\
 0 & 0 & 0 & 0 & 0 & 0 & 0 & 0 & \mathbf{a} \\
\end{array}
\right],
\label{RMA}
\EN
where bold letters are used to 
distinguish the $\mathrm{R}$-matrix elements 
from the Boltzmann weights.

At this point we have the basic ingredients  to
study the possible solutions  
of the Yang-Baxter algebra (\ref{YB}). We shall tackle 
this problem using the following systematic strategy. We start by eliminating the elements
of the $\mathrm{R}$-matrix since the main purpose is the determination of weights fixing the Lax operators.
To this end we search for
suitable functional relations that built a consistent linear system of homogeneous equations
for a particular chosen subset of $\mathrm{R}$-matrix entries. The vanishing of respective determinant is going to
lead us in most cases to polynomials having the anti-symmetrical structure (\ref{fact}). This makes it
possible to define the associated divisors (\ref{factD}) and as a result the freedom of a number of free parameters
$\Lambda_j$. In the situation of functional relations that can not be written directly in the special form (\ref{fact}) we
impose that their polynomials should satisfy 
the anti-symmetric property (\ref{ANTI}). 
This idea is crucial 
to solve very involved functional equations resulting from many nested steps. 
It turns out that we always will be able 
to implement this property at the 
expense of imposing constraints among the free parameters $\Lambda_j$.
As a result either the corresponding polynomial vanishes 
directly or it can be brought into the suitable form (\ref{fact}).
After all that, we still have to perform the intersection of some basic divisors 
which is going to lead us to the main algebraic manifold for the Boltzmann weights. 
As a byproduct we are able to determine the matrix elements
of the $\mathrm{R}$-matrix in terms of few Lax operators weights. In next sections 
we show how to carry out all these steps in practice. 

\section{The Functional Relations}
\label{funrel}

The functional equations constraining the entries of the $\mathrm{R}$-matrix  
and the Boltzmann weights are derived by substituting our
proposals (\ref{LAX},\ref{RMA}) in the Yang-Baxter algebra. We find that 
there exists fifty-four independent equations which are best subdivided
in terms of their number of distinct terms. In Table \ref{TAB} we summarize
this classification which ranges from relations having only two terms to 
those with the maximum number of five elements. 
\begin{table}[ht]
\begin{center}
\begin{tabular}{|c|c|}
\hline
Number of Equations & Number of Terms \\ \hline
2 & two \\ \hline
15 & three \\ \hline
25 & four \\ \hline
12 & five \\ \hline
\end{tabular}
\caption{The number of distinct functional relations versus their
respective number of terms.} \label{TAB}
\end{center}
\end{table}

In general, the technical difficulties in dealing with the solution of the
functional equations increase with their number of terms due to the
presence of many free variables. However, we shall see that the analysis of the
simplest relations with two terms will result in a reasonable decrease of the number of 
independent equations having three elements. This simplification is important to make possible
the solution of such functional equations by the elimination method. In addition to that we will need 
to analyze only a small number of equations with four terms, as compared to the available 
set of Table \ref{TAB},
to decide about the integrability of
the vertex model we have started with. After this study we are left to verify 
that all the remaining polynomial
equations coming from the Yang-Baxter relation are satisfied. It turns out that
this task can be performed algebraically with the help of computer algebra system.

\subsection{Two Terms Equations}

The functional equations having two terms are given as follows,
\begin{eqnarray}
\label{twoeq}
&& \mathbf{c} ({\bar{d}}^{'} d^{''}-d^{'} {\bar{d}}^{''} ) =0,  \nonumber \\
&& (\mathbf{\bar{d}} d^{'}-\mathbf{d} {\bar{d}}^{'} ) c^{''}=0.  
\end{eqnarray}

We see that the above equations are already in the  
convenient form (\ref{fact}) since we are disregarding possible
solutions with zero weights and 
$\mathrm{R}$-matrix amplitudes. Clearly, the corresponding 
divisors fix the following ratios among
the variables,
\EQ
\frac{\mathbf{\bar{d}}}{\mathbf{d}}=
\frac{\bar{d}}{d} = \Lambda_0.
\label{INV0}
\EN

Here we are tacitly assuming that $\Lambda_0 \neq 0$ since otherwise 
we would have to set the weight $\bar{d}$  to zero. Our
main interest is to consider genuine nineteen vertex models and thus
all the Lax operator  weights must be non-null. 
We now turn to investigate more complicated functional relations
having three and four terms.

\subsection{Three Terms Equations}
\label{secthree}

By considering  the ratios (\ref{INV0}) 
the number of independent relations 
with three elements decrease from the original fifteen 
to only nine functional equations. Their
expressions are given by, 
\begin{eqnarray}
\label{threeq1}
&& \mathbf{a} c^{'} a^{''}- \mathbf{\bar{b}} c^{'} b^{''} -\mathbf{c} a^{'} c^{''}=0, \\
\label{threeq2}
&& \mathbf{c} c^{'} b^{''}+ \mathbf{b} a^{'} c^{''} -\mathbf{a} b^{'} c^{''}=0, \\
\label{threeq3}
&& \mathbf{a} d^{'} b^{''}- \mathbf{c} b^{'} d^{''} -\mathbf{\bar{b}} d^{'} f^{''}=0, \\
\label{threeq4}
&& \mathbf{b} b^{'} d^{''}- \mathbf{a} f^{'} d^{''} +\mathbf{c} d^{'} f^{''}=0, \\
\label{threeq5}
&& \mathbf{\bar{b}}  d^{'} a^{''}- \mathbf{f} d^{'} b^{''} -\mathbf{d} {\bar{b}}^{'} c^{''}=0, \\
\label{threeq6}
&& \mathbf{d} \bar{d}^{'} {\bar{b}}^{''}+ \mathbf{f} b^{'} c^{''} -\mathbf{b} f^{'} c^{''}=0, \\
\label{threeq7}
&& \mathbf{c} {\bar{b}}^{'} a^{''}- \mathbf{c} a^{'} {\bar{b}}^{''} -\mathbf{\bar{b}} c^{'} c^{''}=0, \\
\label{threeq8}
&& \mathbf{c} f^{'} {\bar{b}}^{''}- \mathbf{b} \bar{d}^{'} d^{''} -\mathbf{c} {\bar{b}}^{'} f^{''}=0, \\
\label{threeq9}
&& \mathbf{d} f^{'} a^{''}- \mathbf{d} {\bar{b}}^{'} {\bar{b}}^{''} -\mathbf{f} d^{'} c^{''}=0. 
\end{eqnarray}

We first observe that Eqs.(\ref{threeq1}-\ref{threeq9}) 
are not invariant when we exchange 
the $\mathrm{R}$-matrix elements with the
corresponding double primed Lax operator weights. 
The main reason for the absence of this invariance 
is because we are assuming broken $\mathrm{PT}$
symmetry that is $b \neq \bar{b}$ and 
$\mathbf{b} \neq \mathbf{\bar{b}}$.
Therefore, there exists a concrete possibility that
the $\mathrm{R}$-matrix and the Lax operators 
may be sited in two distinct algebraic varieties that are not isomorphic.
This is already an indication we have some chance to obtain
an integrable vertex model whose weights are not trigonometric.

We now consider the solution of Eqs.(\ref{threeq1}-\ref{threeq9}) as a system of
homogeneous relations where 
the unknowns are the $\mathrm{R}$-matrix
entries $\mathbf{a}$, $\mathbf{b}$, $\mathbf{\bar{b}}$, $\mathbf{c}$, $\mathbf{d}$ and $\mathbf{f}$. We have
a number of possibilities of selecting six out of nine equations
to construct a consistent linear system for these variables. 
For a given system to have a non trivial solution 
the determinant of its coefficients
depending on the weights of the Lax operators weights must vanish.
We can for example choose Eqs.(\ref{threeq1}-\ref{threeq6}) 
and the corresponding determinant can be written in the following
form,
\EQ
\left[ ({b^{'}}^2-a^{'} f^{'})c^{''} d^{''}-({b^{''}}^2-a^{''} f^{''})c^{'} d^{'} \right] 
\left[ a^{'} d^{'} c^{''} f^{''} -b^{''} d^{''} b^{'} c^{'}\right] 
\left[ \Lambda_0 {d^{'}}^2 b^{''} {\bar{b}}^{''} -{c^{''}}^2 b^{'} {\bar{b}}^{'} \right]=0.
\label{DET}
\EN

From Eq.(\ref{DET}) we see that in principle we have three possible
branches to be analyzed depending on the factor we choose to vanish.
However, by analyzing other
possible systems of six equations we noted that 
the common factors shared by their determinants are only the
first two terms of Eq.(\ref{DET}). As examples of  
alternative choices of systems we would like to mention those
built up from either Eqs.(\ref{threeq1}-\ref{threeq5},\ref{threeq9}) or  
Eqs.(\ref{threeq1}-\ref{threeq4},\ref{threeq6},\ref{threeq9}). In any of theses
cases the form of the third factor always changes and thus it
plays the role of an extraneous term. From now on we shall disregard
the last factor in Eq.(\ref{DET}) as feasible branch.

We further notice that only the first factor of Eq.(\ref{DET})
has the suitable polynomial form (\ref{fact}). In fact,
the second term of Eq.(\ref{DET}) clearly does not vanishes as  
${\mathbf{w}}^{'} \rightarrow {\mathbf{w}}^{''}$. This is not an impediment
to define an analogous of a divisor but this will leads us to distinct components 
for each of the weights labels. 
At this point we can not discard the second factor as a possible branch since
we have not yet studied the full properties of the linear system.  A more detailed analysis
shall reveal us that such apparent asymmetry of the second factor divisor disappears.

\subsubsection{Main Branch}
\label{threemain}

This branch is defined by imposing that the first polynomial factor of Eq.(\ref{DET}) is zero.
It follows that the corresponding divisor is, 
\EQ
\frac{{b}^2-a f}{c d}= \Lambda_1.
\label{INV1}
\EN

We have now the necessary condition to solve
Eqs.(\ref{threeq1}-\ref{threeq6}) by linear elimination of the variables
$\mathbf{a}$, $\mathbf{b}$, $\mathbf{\bar{b}}$, 
$\mathbf{c}$, $\mathbf{d}$ and $\mathbf{f}$.
At this stage it is sufficient to present their expressions
in a nested form since we have not yet solved the full system
of nine equations. By using Eq.(\ref{INV1}) for double primed labels we
find that these $\mathrm{R}$-matrix entries can be written as,
\begin{eqnarray}
\label{eqv0a}
&&\frac{\mathbf{a}}{\mathbf{c}}=(b^{'} c^{'} b^{''} d^{''}-a^{'} d^{'} c^{''} f^{''} 
)/( \Lambda_1 c^{'} d^{'} c^{''} d^{''}),\\
\label{eqv0bbar}
&&\frac{\mathbf{\bar{b}}}{\mathbf{c}}=( b^{'} c^{'} a^{''} d^{''}-a^{'} d^{'} b^{''} c^{''} 
)/(\Lambda_1 c^{'} d^{'} c^{''} d^{''}),\\
\label{eqv0b}
&&\frac{\mathbf{b}}{\mathbf{c}}=(\frac{\mathbf{a}}{\mathbf{c}} b^{'} c^{''}-c^{'} b^{''} )/( a^{'} c^{''}),\\
\label{eqv0d}
&&\frac{\mathbf{d}}{\mathbf{c}}=\left [ d^{'} c^{''} (\frac{\mathbf{\bar{b}}}{\mathbf{c}} b^{'} a^{''}-
\frac{\mathbf{b}}{\mathbf{c}} f^{'} b^{''}) 
\right ]/\left [ b^{'} {\bar{b}}^{'} {c^{''}}^2-\Lambda_0 {d^{'}}^2 b^{''} {\bar{b}}^{''} \right ],\\
\label{eqv0f}
&&\frac{\mathbf{f}}{\mathbf{c}}=\left [ \frac{\mathbf{b}}{\mathbf{c}} {\bar{b}}^{'} f^{'} {c^{''}}^2 
-\Lambda_0 \frac{\mathbf{\bar{b}}}{\mathbf{c}} {d^{'}}^2 a^{''} \bar{b}^{''} \right ]/ \left [ b^{'} {\bar{b}}^{'} {c^{''}}^2
-\Lambda_0 {d^{'}}^2 
b^{''} {\bar{b}}^{''} \right ],
\end{eqnarray}
where $\mathbf{c}$ is an overall normalization.

We see that Eqs(\ref{eqv0a},\ref{eqv0bbar}) 
are singular when the 
parameter $\Lambda_1$ is zero 
and this means that here we have to  assume $\Lambda_1 \neq 0$. 
As we shall see 
the special case $\Lambda_1=0$ will be covered by the branch related to the 
second polynomial factor of the determinant (\ref{DET}). Let us now begin the analysis of the
remaining three relations considering 
first Eq.(\ref{threeq7}). After substituting the expression for 
the ratio $\mathbf{\bar{b}}/\mathbf{c}$ (\ref{eqv0bbar})
in Eq.(\ref{threeq7}) we 
easily find that it
becomes proportional to the 
polynomial, 
\EQ
(b^{'} c^{'} -\Lambda_1 {\bar{b}}^{'} d^{'})  a^{''} d^{''} -
(b^{''} c^{''} -\Lambda_1 {\bar{b}}^{''} d^{''}) a^{'} d^{'} =0,
\EN
which again has the form (\ref{fact}) and the associated divisor is, 
\EQ
\frac{b c-\Lambda_1 {\bar{b}} d}{a 
d}= \Lambda_2.
\label{INV2}
\EN

We next consider the solution of Eq.(\ref{threeq8}). We first observe that we
have already been able to reduce the number of independent variables by two
weights. 
In fact, considering Eqs.(\ref{INV1},\ref{INV2}) it is not difficult 
to resolve the weights
$f$ and $d$
in terms of the remaining variables $a$, $b$, $\bar{b}$
and $c$. By using this information together with the expression for the ratio
$\mathbf{b}/\mathbf{c}$ (\ref{eqv0b}) and after few simplifications we find that 
Eq.(\ref{threeq8}) can be expressed as follows,
\begin{eqnarray}
&& \left [\Lambda_2 {a^{'}}^2 {\bar{b}}^{'} +\Lambda_1 a^{'} {\bar {b^{'}}^2} -\frac{\Lambda_0}{\Lambda_1}a^{'} {b^{'}}^2 \right ]
\left [ \Lambda_2 a^{''} {{b}^{''}}^2 +\Lambda_1 {b^{''}}^2 {\bar{b}}^{''} -\Lambda_1 b^{''} {c^{''}}^2 \right ] \nonumber \\
&-& \left [\Lambda_2 {a^{''}}^2 {\bar{b}}^{''} +\Lambda_1 a^{''} {\bar {b^{''}}^2} -\frac{\Lambda_0}{\Lambda_1}a^{''} {b^{''}}^2 \right ]
\left [ \Lambda_2 a^{'} {{b}^{'}}^2 +\Lambda_1 {b^{'}}^2 {\bar{b}}^{'} -\Lambda_1 b^{'} {c^{'}}^2 \right ]. \nonumber \\
\label{polINV3}
\end{eqnarray}

Notice that the polynomial (\ref{polINV3}) is already written in the form (\ref{fact}) . Its dependence on 
the parameter $\Lambda_0$ can be re-scaled by means of the following transformation,
\EQ
\Lambda_1={\tilde{\Lambda}_1} \sqrt{\Lambda_0}~~\mathrm{and}~~
\Lambda_2={\tilde{\Lambda}_2} \sqrt{\Lambda_0},
\label{esca}
\EN
and now Eq.(\ref{polINV3}) becomes only dependent on ${\tilde{\Lambda}_1}$ and ${\tilde{\Lambda}_2}$. This re-scaling of parameters
is ultimately due to the freedom of implementing a gauge transformation on the weights
$d$ and $\bar{d}$.
Taking into account Eq.(\ref{esca}) we find that the respective divisor associated to the
polynomial (\ref{polINV3}) is given by,
\EQ
\frac{a \left [b^2-{\tilde{\Lambda}_1} {\tilde{\Lambda}_2} a \bar{b} -{\tilde{\Lambda}_1}^2 {\bar{b}}^2 \right ]}
{b \left[ {\tilde{\Lambda}_2} a {b} +{\tilde{\Lambda}_1} b \bar{b} -{\tilde{\Lambda}_1} c^2 \right]} 
=\Lambda_3.
\label{INV3}
\EN

We have now reached a point in which only Eq.(\ref{threeq9}) remains to be solved.
As before we would like to use the last divisor (\ref{INV3}) to eliminate one further
weight and thus reducing the number of degrees of freedom. We observe however
that the divisor (\ref{INV3}) does not provide us the means to eliminate any 
weight in a linear way. This difficult can be circumvented once we analyze
the polynomial associated to Eq.(\ref{threeq9}) and notice that it depends 
on the weight $c$ only through even powers.  
This means that we can use the fact that divisor (\ref{INV3})
has a quadratic dependence on the weight $c$ to systematically  eliminate this
weight from Eq.(\ref{threeq9}). In Appendix A we explain how this operation can
be implemented within the Mathematica algebraic computer system.
As a result of this procedure we find 
an involved bihomogeneous polynomial with bidegree (4,4)
which unfortunately 
can not be brought into the convenient form (\ref{fact}). At this stage 
of the analysis the requirement that the polynomials 
must satisfy the anti-symmetric property (\ref{ANTI}) 
becomes decisive to make further progress. 
By imposing this property we find that it can indeed be fulfilled provided that the so far
free parameters are constrained by the following simple relation,
\EQ
\Lambda_3 {\tilde{\Lambda}_2}-{\tilde{\Lambda}_1}^2-1=0.
\label{constra1}
\EN

After using the condition (\ref{constra1}) we find that a large number of terms of
the resulting polynomial coming from Eq.(\ref{threeq9}) are magically canceled out.  
Thanks to this simplification  we are able to write the 
resulting polynomial in the appropriate form (\ref{fact}). In what follows
we shall present the respective divisor since from it one can easily 
recover the associated polynomial. The expression for the divisor is given by,
\EQ
\frac{{\tilde{\Lambda}_2} a^2 ({\tilde{\Lambda}_2}^2 \bar{b}^2 -\Lambda_3^2 b^2) +{\tilde{\Lambda}_1} {\tilde{\Lambda}_2} \bar{b}^3 (2 {\tilde{\Lambda}_2} a +{\tilde{\Lambda}_1} \bar{b}) +
b^2 (\Lambda_3^3 \bar{b}^2-{\tilde{\Lambda}_2} b^2)}{\bar{b} b^2 ({\tilde{\Lambda}_1} \bar{b}+{\tilde{\Lambda}_2} a)}
=\Lambda_4.
\label{INV4}
\EN

This completes the solution of the nine functional relations for this branch. Up to this point
the effective integrable manifold of the weights should be given by the intersection of the last 
two divisors (\ref{INV3},\ref{INV4}). In addition to that we have four free parameters at our disposal since so
far we just have the constrain (\ref{constra1}).
These divisors give rise to two projective surfaces $\in \mathbb{CP}^3$ and is generally expected that 
their intersection will lead us to an algebraic spatial curve. Since some of the free parameters are going
to be fixed later on we shall postpone the analysis of the intersection until the very end.

\subsubsection{Special Branch}
\label{threeespecial}

This branch is defined by setting the second term of Eq.(\ref{DET}) to zero and it
is related to the particular value $\Lambda_1=0$ excluded in the main branch. This can be
seen by first noticing that  
the $\mathrm{R}$-matrix
ratio $\mathbf{a}/\mathbf{c}$ (\ref{eqv0a}) is proportional to the 
second factor of the determinant (\ref{DET}). 
For this ratio to be no null we should have another zero on the
denominator such that the value of the ratio becomes indeterminate. 
Inspecting Eq.(\ref{eqv0a}) we note that the only option is to set 
$\Lambda_1=0$ and as result it follows that
the weight $f$ must be fixed by the expression,
\EQ
f=\frac{b^2}{a}.
\label{DIV1}
\EN

We now  observe that the numerator 
of Eq.(\ref{eqv0bbar}) must vanish otherwise we would have a divergence on 
the ratio $\mathbf{\bar{b}}/\mathbf{c}$. From this condition it follows a polynomial 
relation of the form (\ref{fact}), namely
\EQ
b^{'} c^{'} a^{''} d^{''}-b^{''} c^{''} a^{'} d^{'}=0,
\EN
whose corresponding divisor is,
\EQ
\frac{b c}{a d}=\Lambda_2.
\label{DIV2}
\EN

The next step is to evaluate the indeterminacy of the above mentioned 
$\mathrm{R}$-matrix ratios. This is fortunately done with
the help of Eqs.(\ref{threeq1},\ref{threeq7}) and the final expressions for such ratios are, 
\begin{eqnarray}
&& \mathbf{a}/\mathbf{c}= [\bar{b}^{'} a^{''} b^{''} -a^{'} b^{''} {\bar{b}^{''}} +a^{'} {c^{''}}^2]/(c^{'} c^{''} a^{''}), \\
&& \mathbf{\bar{b}}/\mathbf{c}= ({\bar{b}^{'}} a^{''} -a^{'} {\bar{b}^{''}})/(c^{'} c^{''}),
\end{eqnarray}
while the other $\mathrm{R}$-matrix entries can again be 
computed from Eqs.(\ref{eqv0b}-\ref{eqv0f}).

At this stage the only relations 
that remain to be solved 
are Eqs.(\ref{threeq8},\ref{threeq9}).
The technical details entering their solution are fairly parallel
to those already explained in the previous subsection for the main
branch. In what follows we shall therefore present only the final
results for the corresponding divisors. We find that the divisor
associated to Eq.(\ref{threeq8}) has the following form,
\EQ
\frac{{\tilde{\Lambda}_2} a^2\bar{b}}{b c^2- b^2 \bar{b}}=\Lambda_3,
\label{DIV3}
\EN
where $\Lambda_2$ has been re-scaled as in Eq.(\ref{esca}) and together 
with the parameter $\Lambda_3$ they satisfy the constraint,
\EQ
\Lambda_3 {\tilde{\Lambda}_2}-1=0.
\label{conLAMBDA3ES}
\EN

Finally, the divisor associated to the 
anti-symmetrical polynomial 
derived from Eq.(\ref{threeq9}) is given by,
\EQ
\frac{{\tilde{\Lambda}_2} a^2 ({\tilde{\Lambda}_2}^2 \bar{b}^2 -\Lambda_3^2 b^2) +
b^2 (\Lambda_3^3 \bar{b}^2-{\tilde{\Lambda}_2} b^2)}{{\tilde{\Lambda}_2} \bar{b} b^2 }
=\Lambda_4.
\label{DIV4}
\EN

We note that the divisors (\ref{DIV1},\ref{DIV2},\ref{DIV4}) can 
be directly obtained from those derived for 
the main branch by substituting ${\tilde{\Lambda}_1}=0$
in Eqs.(\ref{INV1},\ref{INV2},\ref{INV4}), respectively. The same observation
for the remaining divisor (\ref{DIV3}) is more subtle since the zero
order in ${\tilde{\Lambda}_1}$ of the corresponding main branch divisor (\ref{INV3})
is trivial. In spite of that we can recover the divisor (\ref{DIV3})
by rewriting Eq.(\ref{INV3}) in powers of 
the parameter ${\tilde{\Lambda}_1}$. By setting the coefficient proportional to
${\tilde{\Lambda}_1}$ to zero we then easily obtain the divisor (\ref{INV3}).
In practice, since this limit is somewhat delicate we shall 
consider these two branches separately.

\subsection{Four Terms Equations}
\label{secfour}

The Boltzmann weights $h$, $\bar{h}$ and $g$ begin to emerge only in the functional equations having four
distinct terms. Even after using the divisor (\ref{INV0}) we find that the total number of relations with four
terms is still considerable. Altogether we have twenty-two functional equations which is a high number to approach
the problem by the standard elimination method. As we shall see however we need to solve only ten functional
relations to decide on the integrability of the vertex model. 
The reason for this simplification is that few of the 
equations with four terms have as unknowns the
$\mathrm{R}$-matrix entries $\mathbf{a}$, $\mathbf{b}$ and $\mathbf{\bar{b}}$ which 
have already been fixed. For sake of consistent of the elimination procedure we are
then forced to make linear combinations between a small subset of four terms functional
equations and two particular three terms relations. Remarkably enough this simple
reasoning is able to determine 
the structure of the remaining weights $h$, $\bar{h}$ and $g$.  We shall first detail
how this procedure works for the weight $g$.

\subsubsection{The weight $g$}
\label{subweig}
We have three functional equations depending uniquely on the weight $g$ and on some variables 
that have been previously determined. Their expressions are given by,
\begin{eqnarray}
\label{foureqg1}
&& \mathbf{c} g^{'} b^{''} +\mathbf{b} c^{'} c^{''} -\mathbf{\bar{b}}{\bar{d}}^{'} d^{''} -\mathbf{c} b^{'} g^{''}=0, \\
\label{foureqg2}
&& \mathbf{g} d^{'} {\bar{b}}^{''} +\mathbf{d} b^{'} c^{''} -\mathbf{c}{\bar{b}}^{'} d^{''} -\mathbf{b} d^{'} g^{''}=0, \\
\label{foureqg3}
&& \mathbf{g} {\bar{b}}^{'} c^{''} +\mathbf{d} {\bar{d}}^{'} b^{''} -\mathbf{c} c^{'} {\bar{b}}^{''} -\mathbf{\bar{b}} g^{'} c^{''}=0. 
\end{eqnarray}

We see that Eqs.(\ref{foureqg1}-\ref{foureqg3}) have five unknowns and therefore we need more two equations to
build up a consistent homogeneous linear system. These extra relations should be searched among those solved in 
subsection (\ref{threemain})
since the above equations depend on the 
weights $\mathbf{b}$, $\mathbf{\bar{b}}$ and $\mathbf{c}$. Direct inspection of Eqs.(\ref{threeq1}-\ref{threeq9})  
reveals us that this choice is remarkably unique once we want to keep the minimal number of five unknowns. These
suitable relations turn out to be  Eqs.(\ref{threeq7},\ref{threeq8}). Now, by setting
the determinant of equations (\ref{threeq7},\ref{threeq8},\ref{foureqg1}-\ref{foureqg3}) equal to zero we find
that it can be factorized as,
\begin{eqnarray}
&& \left [\Lambda_0 (b^{'} c^{'} d^{'} c^{''} d^{''} g^{''} -b^{''} c^{''} d^{''} c^{'} d^{'} g^{'})
+\Lambda_0^2(\bar{b}^{'} {d^{'}}^2 a^{''} {d^{''}}^2 -\bar{b}^{''} {d^{''}}^2 a^{'} {d^{'}}^2)  \right. \nonumber \\
&& \left. +\bar{b}^{'} {c^{'}}^2 {c^{''}}^2 f^{''} -\bar{b}^{''} {c^{''}}^2 {c^{'}}^2 f^{'}  \right ] 
\left[ b^{'} \bar{b}^{'} {c^{''}}^2 -\Lambda_0 {d^{'}}^2 b^{''} \bar{b}^{''} \right]=0.
\label{DET1}
\end{eqnarray}

The determinant (\ref{DET1}) must vanish through the first factor since the second one is exactly
the extraneous term that has been discarded before. We see that the first factor of Eq.(\ref{DET1}) is anti-symmetrical
on the exchange of weights labels and  taking into account 
the divisors (\ref{INV1},\ref{INV2},\ref{INV3}) we are indeed able to rewrite it
in the appropriate form (\ref{fact}). The resulting polynomial depends only on the
weights $a$, $b$ ,$\bar{b}$, $g$ and the corresponding divisor is given by,
\EQ
\frac{{\tilde{\Lambda}_2} \Lambda_3 b({\tilde{\Lambda}_2} a+{\tilde{\Lambda}_1} \bar{b})g -
\bar{b}[\Lambda_3 b^2 +{\tilde{\Lambda}_2} ({\tilde{\Lambda}_2} a +{\tilde{\Lambda}_1} \bar{b})^2]}{b^2 ({\tilde{\Lambda}_2}a +{\tilde{\Lambda}_1} \bar{b})} =\Lambda_5
\label{INV5}
\EN

Let us consider the solution of the remaining equations (\ref{foureqg2},\ref{foureqg3}). We can 
easily solve one of these relations fixing the value of the 
$\mathrm{R}$-matrix element $\mathbf{g}$. For instance from Eq.(\ref{foureqg2}) we find, 
\EQ
\frac{\mathbf{g}}{\mathbf{c}}= \left [ 
\frac{\mathbf{b}}{\mathbf{c}}d^{'} g^{''} +\bar{b}^{'} d^{''} 
-\frac{\mathbf{d}}{\mathbf{c}}b^{'} c^{''} 
\right ]/(d^{'} \bar{b}^{''})
\EN
where the ratios 
$\mathbf{b}/\mathbf{c}$ and $\mathbf{d}/\mathbf{c}$ have already 
been determined by Eqs.(\ref{eqv0b},\ref{eqv0d}). 

The last equation (\ref{foureqg3}) can now be rewritten only in terms of the  
weights $a$, $b$ and $\bar{b}$ once we use the help of the 
divisors (\ref{INV1},\ref{INV2},\ref{INV3},\ref{INV5}).  
After some simplifications we find that 
Eq.(\ref{foureqg3}) becomes proportional to the following expression,   
\begin{eqnarray}
&& \left [ \Lambda_5-{\tilde{\Lambda}_1}(\Lambda_3+{\tilde{\Lambda}_2}) \right] \left[ {b^{''}}^2({\tilde{\Lambda}_2} \Lambda_3^2 {a^{'}}^2 +{\tilde{\Lambda}_2} {b^{'}}^2
+{\tilde{\Lambda}_2} \Lambda_5 a^{'} \bar{b}^{'} +({\tilde{\Lambda}_1} \Lambda_5-\Lambda_3^3) {\bar {b^{'}}^2}) + \right. \nonumber \\
&& \left. {\tilde{\Lambda}_2} (\Lambda_3-{\tilde{\Lambda}_2}) a^{''} \bar{b}^{''}({\tilde{\Lambda}_2} a^{'} \bar{b}^{'} +{\tilde{\Lambda}_1} {\bar {b^{'}}^2} )
-{\tilde{\Lambda}_2} {\bar {b^{''}}^2} ({\tilde{\Lambda}_2} \Lambda_3 {a^{'}}^2 +{\tilde{\Lambda}_1} (\Lambda_3 +{\tilde{\Lambda}_2}) a^{'} \bar{b}^{'} +{\tilde{\Lambda}_1}^2 
{\bar {b^{'}}^2}
) \right ].
\label{POLG}
\end{eqnarray}

The polynomial (\ref{POLG}) is very far from satisfying 
the anti-symmetrical property (\ref{ANTI}) but nevertheless it is proportional
to a combination of free parameters.  We then are able to solve 
Eq.(\ref{foureqg3}) by choosing that its first factor vanishes and
as result the parameter $\Lambda_5$ becomes fixed by,
\EQ
\Lambda_5={\tilde{\Lambda}_1}(\Lambda_3+{\tilde{\Lambda}_2})
\label{LAM5}
\EN

We finally remark that though the above discussion  has been detailed for
main branch there is no difficulty to repeat the same analysis 
in the case of the special branch. It turns out that for the
special branch we simply have to set ${\tilde{\Lambda}_1}=0$ in the final
results (\ref{INV5},\ref{LAM5}).

\subsubsection{Weights $h$ and $\bar{h}$ }

Out of thirteen functional equations involving the weights $h$ and $\bar{h}$ we shall
need only seven of them to determine both these weights and the $\mathrm{R}$-matrix
elements $\mathbf{h}$ and $\mathbf{\bar{h}}$. The expressions of such basic relations are,

\begin{eqnarray}
\label{foureqh1}
&& \mathbf{c} h^{'} \bar{b}^{''} +\mathbf{\bar{b}} c^{'} c^{''} -\mathbf{b}d^{'} d^{''} -\mathbf{c} \bar{b}^{'} h^{''}=0, \\
\label{foureqh2}
&& \mathbf{c} \bar{h}^{'} \bar{b}^{''} +\mathbf{\bar{b}} c^{'} c^{''} -\mathbf{b}\bar{d}^{'} \bar{d}^{''} 
-\mathbf{c} \bar{b}^{'} \bar{h}^{''}=0, \\
\label{foureqh3}
&& \mathbf{h} b^{'} {c}^{''} -\mathbf{b} h^{'} c^{''} -\mathbf{c}c^{'} b^{''} +\mathbf{d} {d}^{'} \bar{b}^{''}=0, \\
\label{foureqh4}
&& \mathbf{\bar{h}} b^{'} {c}^{''} -\mathbf{b} \bar{h}^{'} c^{''} -\mathbf{c}c^{'} b^{''} +\mathbf{\bar{d}} {\bar{d}}^{'} \bar{b}^{''}=0, \\
\label{foureqh5}
&& \mathbf{h} \bar{d}^{'} {b}^{''} +\mathbf{d} \bar{b}^{'} c^{''} -\mathbf{c}b^{'} \bar{d}^{''} 
-\mathbf{\bar{b}} {d}^{'} \bar{h}^{''}=0, \\
\label{foureqh6}
&& \mathbf{\bar{h}} d^{'} {b}^{''} +\mathbf{\bar{d}} \bar{b}^{'} c^{''} -\mathbf{c}b^{'} d^{''} 
-\mathbf{\bar{b}} {\bar{d}}^{'} {h}^{''}=0, \\
\label{foureqh7}
&& \mathbf{a} h^{'} {a}^{''} -\mathbf{d} c^{'} d^{''} -\mathbf{f} h^{'} f^{''} -\mathbf{h} {a}^{'} {h}^{''}=0. 
\end{eqnarray}

As before we observe that Eqs.(\ref{foureqh1},\ref{foureqh2}) have 
as unknowns the $\mathrm{R}$-matrix elements $\mathbf{a}$, 
$\mathbf{b}$ and $\mathbf{\bar{b}}$ and these equations can again be solved by 
making linear combinations
with Eqs.(\ref{threeq7},\ref{threeq8}). By requiring that the determinant of 
Eqs.(\ref{threeq7},\ref{threeq8},\ref{foureqh1}) vanishes we obtain,
\EQ
c^{'}d^{'}c^{''}d^{''} \left [ \bar{b}^{'} (h^{''}-a^{''}-f^{''}/\Lambda_0) -
\bar{b}^{''}(h^{'}-a^{'}-f^{'}/\Lambda_0) \right]=0,
\EN
while the vanishing of the one made by system of 
equations (\ref{threeq7},\ref{threeq8},\ref{foureqh2}) is,
\EQ
c^{'}d^{'}c^{''}d^{''} \left [ \bar{b}^{'} (\bar{h}^{''}-a^{''}-\Lambda_0 f^{''}) -
\bar{b}^{''}(\bar{h}^{'}-a^{'}-\Lambda_0 f^{'}) \right]=0.
\EN

The second factors of the above relations give rise to anti-symmetrical polynomials 
of the form (\ref{fact}) and their corresponding divisors determine
the expressions for the weights $h$ and $\bar{h}$, 
\EQ
\frac{h-a-f/\Lambda_0}{\bar{b}}=\Lambda_6~~~\mathrm{and}~~~
\frac{\bar{h}-a-\Lambda_0f}{\bar{b}}=\bar{\Lambda}_6
\label{INV6}
\EN

We next note that from Eqs.(\ref{foureqh3},\ref{foureqh4}) we can retrieve the
$\mathrm{R}$-matrix entries $\mathbf{h}$ and $\mathbf{\bar{h}}$.
We further observe
that the last two terms of these equations are also present in the previously solved
three terms relations (\ref{threeq2},\ref{threeq6}). This makes it possible to eliminate
the terms $\mathbf{c} c^{'} b^{''}$ and $\mathbf{d}\bar{d}^{'} \bar{b}^{''}$ 
of Eqs.(\ref{foureqh3},\ref{foureqh4}) and  
after using the divisors (\ref{INV6}) for the single primed labels 
we find that the ratios $\mathbf{h}/\mathbf{c}$ 
$\mathbf{\bar{h}}/\mathbf{c}$  are given by,
\EQ
\frac{\mathbf{h}}{\mathbf{c}}= \frac{\mathbf{a}}{\mathbf{c}}+(\frac{\mathbf{f}}{\mathbf{c}})/\Lambda_0+ 
\Lambda_6 \frac{\bar{b}^{'}}{b^{'}} \frac{\mathbf{b}}{\mathbf{c}},~~~\mathrm{and}~~~
\frac{\mathbf{\bar{h}}}{\mathbf{c}}= \frac{\mathbf{a}}{\mathbf{c}}+\Lambda_0(\frac{\mathbf{f}}{\mathbf{c}})+ 
\bar{\Lambda}_6 \frac{\bar{b}^{'}}{b^{'}} \frac{\mathbf{b}}{\mathbf{c}}
\label{hzero}
\EN

We now have the basic 
ingredients to tackle the solution of Eqs.(\ref{foureqh5},\ref{foureqh6}). Considering the above results 
altogether as well as the previous expressions for the ratios $\mathbf{d}/\mathbf{c}$ and
$\mathbf{\bar{b}}/\mathbf{c}$ we are able to write these equations solely
in terms of the weights $a$, $b$ and $\bar{b}$. 
We find that the vanishing of Eqs.(\ref{foureqh5},\ref{foureqh6}) are equivalent to the
the following identities,
\EQ
\label{POLINVh1}
\Lambda_6 \Lambda_0 \bar{b}^{'} \left [ {\tilde{\Lambda}_2}({b^{'}}^2 a^{''} \bar{b}^{''} -{b^{''}}^2 a^{'} \bar{b}^{'}) 
+{\tilde{\Lambda}_1}({b^{'}}^2 {\bar{b^{''}}^2} -{{b}^{''}}^2 {\bar{b^{'}}^2} ) \right ] 
+\bar{\Lambda}_6 \Lambda_3 {b^{'}}^2 \bar{b}^{''}(a^{'} \bar{b}^{''}-a^{''} \bar{b}^{'})=0,
\EN
\EQ
\label{POLINVh2}
\bar{\Lambda}_6 \bar{b}^{'} \left [ {\tilde{\Lambda}_2}({b^{'}}^2 a^{''} \bar{b}^{''} -{b^{''}}^2 a^{'} \bar{b}^{'}) 
+{\tilde{\Lambda}_1}({b^{'}}^2 {\bar{b^{''}}^2} -{{b}^{''}}^2 {\bar{b^{'}}^2} ) \right ] 
+\Lambda_6 \Lambda_3 \Lambda_0 {b^{'}}^2 \bar{b}^{''}(a^{'} \bar{b}^{''}-a^{''} \bar{b}^{'})=0.
\EN

We see that Eq.(\ref{POLINVh1},\ref{POLINVh2}) contain common 
anti-symmetrical polynomials which in principle could be set to zero
by means of the corresponding divisors. This possibility imposes at least an extra
constrain on the weights $a$ and $\bar{b}$ in addition to the other two divisors (\ref{INV3},\ref{INV4})
already derived in subsection (\ref{threemain}). The intersection of such three divisors will generically
lead us to zero dimensional manifold consisted 
of finite number of points for the ratios $a/c$, $b/c$ and $\bar{b}/c$ and thus to 
Lax operators without free spectral parameters. From now on we shall discard this kind
of possible ``braid'' solutions of the Yang-Baxter algebra.
However, it is fortunate that Eqs.(\ref{POLINVh1},\ref{POLINVh2}) can vanish without the
definition of any additional divisor  by choosing 
the last two free parameters to be zero, that is, 
\EQ	
\Lambda_6=\bar{\Lambda}_6=0
\label{conLAMBDA6}
\EN

Let us finally consider the solution of Eq.(\ref{foureqh7}). Considering all the
results we have obtained so far we find that Eq.(\ref{foureqh7})  leads us 
once again to deal with a very complicated polynomial
of bidegree (4,4) on the weights $a$, $b$ and $\bar{b}$. To make progress we then
impose the anti-symmetrical property (\ref{ANTI}) in the expectation of further
simplifications. Fortunately, we find that such property can be satisfied 
provided we introduce a new constrain among the parameters, namely
\EQ
\Lambda_4 {\tilde{\Lambda}_1} \Lambda_0+\Lambda_3^2 \left[ {\tilde{\Lambda}_2}(1+\Lambda_0^2)
-\Lambda_3 \Lambda_0) \right ] -{\tilde{\Lambda}_2} \Lambda_0(1-{\tilde{\Lambda}_1}^2)=0.
\label{conLAMBDA4}
\EN

By substituting the constrain (\ref{conLAMBDA4}) 
back into Eq.(\ref{foureqh7}) we find that its non trivial terms become
proportional to the following polynomial,  
\EQ
(\Lambda_0^2-\Lambda_0+1) 
(a^{'} \bar{b}^{''}-a^{''} \bar{b}^{'})
\left [ {\tilde{\Lambda}_1} a^{'} ({b^{'}}^2+{\bar{b^{'}}^2}) 
+\bar{b}^{'} ({\tilde{\Lambda}_2} {a^{'}}^2+\Lambda_3 {b^{'}}^2) \right ] 
\left [ {\tilde{\Lambda}_1} a^{''} ({b^{''}}^2+{\bar{b^{''}}^2}) 
+\bar{b}^{''} ({\tilde{\Lambda}_2} {a^{''}}^2+\Lambda_3 {b^{''}}^2) \right ]. 
\label{POL86}
\EN

Taking into account the above discussion we conclude that Eq.(\ref{foureqh7}) should vanishes 
by imposing that
the first factor of (\ref{POL86}) is zero. As a consequence
the parameter $\Lambda_0^2$ is constrained to take values on
the non trivial third order roots of unity, namely
\EQ
\Lambda_0= \exp(\pm \mathrm{i} \frac{\pi}{3}).
\label{conLAMBDA0}
\EN

Now by considering the above constrain for $\Lambda_0$ back to Eq.(\ref{conLAMBDA4})  we see that
this parameter factorizes leading us to an effective
condition independent of $\Lambda_0$,
\EQ
\Lambda_4 {\tilde{\Lambda}_1} +\Lambda_3^2 ({\tilde{\Lambda}_2}
-\Lambda_3 ) -{\tilde{\Lambda}_2} (1-{\tilde{\Lambda}_1}^2)=0.
\label{conLAMBDA4EF}
\EN

Although the above discussion has been concentrated in the case 
of the main branch the same reasoning applies straightforwardly 
for the special branch. Once again we just have   
to impose ${\tilde{\Lambda}_1}=0$ in the final results and we see that in practice this
becomes relevant only for Eq.(\ref{conLAMBDA4EF}). By substituting
${\tilde{\Lambda}_1}=0$ in this relation and with the help of Eq.(\ref{conLAMBDA3ES})
we conclude that for the special branch the parameter ${\tilde{\Lambda}_2}$ is required to 
satisfy the relation ${\tilde{\Lambda}_2}^4-{\tilde{\Lambda}_2}^2+1=0$. This means that the
for special branch all the parameters have already been fixed with the
exception of $\Lambda_4$. By way of contrast for the main branch 
we have two free parameters and to avoid division by a potential
zero factor we choose them to be ${\tilde{\Lambda}_2}$
and $\Lambda_3$. 
For sake of clearness we have summarized the values
of fixed parameters for both branches in Table \ref{TAB1}. 
\begin{table}[ht]
\begin{center}
\begin{tabular}{|c|c|c|c|c|c|c|c|}
\hline
Parameters & $\Lambda_0$ & ${\tilde{\Lambda}_1}$ & ${\tilde{\Lambda}_2}$ & $\Lambda_3$ & $\Lambda_4$ & $\Lambda_5$ & $\Lambda_6,\bar{\Lambda}_6$ \\ \hline
Main Branch & $ \exp(\pm \mathrm{i} \frac{\pi}{3})$  & $\pm \sqrt{{\tilde{\Lambda}_2} \Lambda_3-1}$ & $\mathrm{Free}$ & $\mathrm{Free} $ 
& $\frac{\Lambda_3^2(\Lambda_3-{\tilde{\Lambda}_2})+{\tilde{\Lambda}_2}(1-{\tilde{\Lambda}_1}^2)}{{\tilde{\Lambda}_1}}$ & ${\tilde{\Lambda}_1}(\Lambda_3+{\tilde{\Lambda}_2})$ & 0 \\ \hline
Special Branch & $ \exp(\pm \mathrm{i} \frac{\pi}{3})$  & 0 & $\pm \Lambda_0^{\pm \frac{1}{2}}$ & ${\tilde{\Lambda}_2}^{-1} $ 
& $\mathrm{Free}$ & 0 & 0 \\ \hline
\end{tabular}
\caption{The values of the parameters associated to the divisors for both branches.} 
\label{TAB1}
\end{center}
\end{table}

We now reach a point in which there are twenty-four functional relations
involving four and five terms that still have to be verified. To avoid overcrowding
this section with a number of additional formulae we have 
presented their explicit expressions in Appendix A. Taking
into account the results obtained so far the
bihomogeneous polynomials associated to Eqs.(\ref{fourextra1}-\ref{fiveextra2}) 
can clearly be expressed
in terms of the weights $a$, $b$, $\bar{b}$ and $c$. 
In the case of the main branch we can use the help of
the divisors (\ref{INV3},\ref{INV4})  
to simplify these polynomials in a systematic way until
we reach the point where the weight $c$ is completely eliminated and
the powers of the weights $\bar{b}$ are at most three. Remarkably enough,
after such algebraic manipulations these twenty-four polynomials
are either zero or become proportional to the factor $\Lambda_0^2-\Lambda_0+1$ and
consequently vanish thanks to the constrain (\ref{conLAMBDA0}). 
Similar reasoning can be implemented
for the special branch by considering now 
the respective divisors (\ref{DIV3},\ref{DIV4}). The technical details  
concerning these simplifications are described in Appendix A including
the explicit computer algebra system code used.
In this way we are able
to verify $\bf{algebraically}$ that the whole Yang-Baxter algebra (\ref{YB}) 
is indeed satisfied for both branches.
We emphasize 
that this verification can be done in rather modest computer
as far as memory is concerned and without the need
to take any a priori numeric values either for the available 
free parameters or for the Boltzmann weights.

\section{The Manifolds Geometry}
\label{geoman}

The purpose of this section is to present the structure of
the main algebraic manifolds governing the integrability of
the previously uncovered vertex models as well as to discuss their
geometric properties.
We shall argue that the weights of the Lax operators can be written
solely in terms of three variables $a$, $b$ and $c$  which are 
constrained by projective plane algebraic curves of genus five for generic
values of the free parameters. Since we will be dealing mostly with
singular manifolds we start by recalling the definition
of singular locus of an irreducible hypersurface 
$S(\omega_0,\hdots,\omega_m) \in \mathbb{CP}^m$. The set of singular 
points of this hypersurface form a closed subvariety $\mathrm{Sing}(S)$ determined by
the zeroes of all the partial 
derivatives of $S(\omega_0,\hdots,\omega_m)$,
\EQ
\label{sing}
\mathrm{Sing}(S)=\{[\omega_{0}:\hdots:\omega_m]
\in \mathbb{CP}^m | \frac{\partial S}{\partial \omega_j}=0,~\mathrm{for}~,j=0,\hdots,m.
\} 
\EN

The analysis of the singularities is going to be helpful to elucidate the intersection problem
of surfaces in the case of the main branch. This study is also  essential for 
the understanding of the geometric
properties of the algebraic curves we shall confront here.

\subsection{The Main Branch}
\label{maincurves}

From the majority of the divisors we are able to extract the weights in terms
of the variables $a$, $b$ and $c$ by
means of linear elimination. The only exception for this branch is the
weight $\bar{b}$ since we have to deal with the intersection 
of the divisors (\ref{INV3},\ref{INV4}) which are both non-linear in this
variable. Before considering this problem  we find 
convenient to perform the following re-scaling on the
parameters ${\tilde{\Lambda}_2}$ and $\Lambda_3$,
\EQ
\label{resca}
{\tilde{\Lambda}_2} ={\tilde{\Lambda}_1} \varepsilon_1 ~~\mathrm{and}~~\Lambda_3={\tilde{\Lambda}_1} \varepsilon_2.
\EN
where $\varepsilon_1$ and $\varepsilon_2$ are now the two free parameters 
associated to the main branch.

Considering the above re-scaling and the parameters values of Table (\ref{TAB1})
we can rewrite the divisors (\ref{INV3},\ref{INV4}) as two projective surfaces in $\mathbb{CP}^3$. 
The polynomial expression associated to divisor (\ref{INV3}) becomes,
\begin{eqnarray}
\label{surfS1}
S_1(a,b,\bar{b},c) &=& a (b^2 +\bar{b}^2) +(\varepsilon_1 a^2 +\varepsilon_2 b^2) \bar{b} -\varepsilon_2 b c^2,
\end{eqnarray}
while the divisor (\ref{INV4}) can be rewritten as,
\begin{eqnarray}
\label{surfS2}
S_2(a,b,\bar{b}) &=& a^2 (\varepsilon_1^2 \bar{b}^2-\varepsilon_2^2 b^2) +(2\varepsilon_1 a +\bar{b}) \bar{b}^3
-(\varepsilon_1 \varepsilon_2 -\varepsilon_2^2-2)b^2 \bar{b}^2  -(\varepsilon_1 \varepsilon_2-1)b^4 \nonumber \\
&-& (\varepsilon_1^2 \varepsilon_2-\varepsilon_1 \varepsilon_2^2-2\varepsilon_1+\varepsilon_2^3) a b^2 \bar{b}.
\end{eqnarray}

The surface $S_1$ is free of singularities for generic values of the parameters $\varepsilon_1$
and $\varepsilon_2$ because there is no solution to Eq.(\ref{sing}) other than the origin. This is
a rational manifold since any nonsingular cubic surface is known to be birational to $\mathbb{CP}^2$ \cite{SHAFA}.
The surface $S_2$ does not depend on the weight $c$ and thus has to be seen as cone over a plane algebraic
curve. By solving Eq.(\ref{sing}) for $S_2$ we find that it contains 
two non coplanar singular lines given by,
\EQ
\label{singS2}
{\bf{l}}_1=[a:0:0:1] ~~\mathrm{and}~~{\bf{l}}_2=[a:0:-\varepsilon_1a:1],
\EN
and for this reason it is expected that $S_2$ should be 
a cone over an elliptic curve \cite{BEAU}. In appendix B
we present the technical details showing that $S_2$ 
indeed defines a curve of genus one when viewed    
in the ring $\mathbb{C}[a,b,\bar{b}]$.

Let us turn to the problem of computing the intersection of the pair of
algebraic surfaces (\ref{surfS1},\ref{surfS2}). From the algebraic geometry theory we know 
that the result of the intersection of two
irreducible surfaces $S_1$ and $S_2$ $\subset \mathbb{CP}^3$ is  
always a finite collection of 
curves \cite{HAR}. An extension of the Bezout's theorem asserts 
that the product of the degrees of $S_1$ and $S_2$ should satisfy the relation,
\EQ
\label{deg}
\mathrm{deg}(S_1) \mathrm{deg}(S_2)= \sum_{\alpha} \mathrm{I}_{\alpha}(S_1,S_2) \mathrm{C}_{\alpha}.
\EN
where 
the sum varies over all the irreducible curves $\mathrm{C}_{\alpha}$ of $S_1 \cap S_2$.
The index $\mathrm{I}_{\alpha}(S_1,S_2)$ is the multiplicity of the intersection of $S_1$ and $S_2$ along
the component $\mathrm{C}_{\alpha}$.  

The first step of the problem is therefore to find out 
whether the intersection is irreducible or it degenerates in the
union of a given number of curves. 
This can be answered by exploring a
basic fact in commutative algebra that assures us that the ideal generated
by the polynomials $S_1$ and $S_2$ can be decomposed in terms of more
elementary ideal components, see \cite{COMU} for mathematical details.
These components are called primary ideals 
whose defining polynomials give rise to the varieties corresponding
precisely to the irreducible curves coming from the intersection $S_1 \cap S_2$.
It is fortunate that the existing algorithms for extracting
information on this primary decomposition have already 
been implemented in
computer algebra systems such as for example Singular \cite{SIN}.
With the help of this software we find that the ideal 
associated to the polynomials (\ref{surfS1},\ref{surfS2}) is in fact reducible  
having three main primary components. Here we are in a favorable 
situation in which two of them sit exactly at the
singular locus of $S_2$ defined by the lines (\ref{singS2}).
By analyzing the factorization of the polynomials (\ref{surfS1},\ref{surfS2})
around $b=0$ we can easily obtain that the respective intersection
indices are 
$\mathrm{I}_{{\bf{l}}_1}(S_1,S_2) =
\mathrm{I}_{{\bf{l}}_2}(S_1,S_2) =2$.  
Considering all these information together
with formula (\ref{deg}) we conclude that the third component 
has to be a spatial curve of degree eight. From the practical 
point of view such third component is the only non trivial
result of the intersection since we are not interested in solutions
having null weights. 

The next step is to map the surface  intersection to a 
degree eight plane curve by hopefully determining 
exactly one of the weights in terms of the curve variables.
To this end the result of the primary decomposition for
the third component is of any help since the respective
basis of the primary ideal is constructed out  of a large number
of polynomials. This task can however be resolved
by means of the following method. We rewrite the
the surfaces expressions as univariate polynomials 
in the variable we want
to eliminate which here we choose to be 
the weight $\bar{b}$, namely 
\begin{eqnarray}
\label{surfS1a}
S_1(\bar{b}) & = & {\bf{v}}_2 \bar{b}^2 +{\bf{v}}_1 \bar{b} +{\bf{v}}_0,  \\
\label{surfS2a}
S_2(\bar{b}) & = & {\bf{u}}_4 \bar{b}^4+{\bf{u}}_3 \bar{b}^3+ {\bf{u}}_2 \bar{b}^2 +{\bf{u}}_1 \bar{b} +{\bf{u}}_0,
\end{eqnarray}
where the coefficients $\bf{v}_j$ and $\bf{u}_j$ depend on the remaining weights 
$a$, $b$ and $c$. They are determined by just matching Eqs.(\ref{surfS1},\ref{surfS2})
to Eqs.(\ref{surfS1a},\ref{surfS2a}) respectively.

We now proceed by making certain linear combinations among
$S_1(\bar{b})$ and $S_2(\bar{b})$ in order to lower the
highest degree in the variable $\bar{b}$. For instance this is
achieved by defining the following new polynomials,
\begin{eqnarray}
\label{lower}
\tilde{S}_1(\bar{b}) & = & \left [ {\bf{u}}_0 S_1(\bar{b}) -{\bf{v}}_0 S_2(\bar{b}) \right]/\bar{b}, \nonumber \\
\tilde{S}_2(\bar{b}) & = & {\bf{u}}_4 \bar{b}^2 S_1(\bar{b}) -{\bf{v}}_2 S_2(\bar{b}).
\end{eqnarray}

Because both $\tilde{S}_1(\bar{b})$ and $\tilde{S}_2(\bar{b})$ are 
algebraic combinations of
the starting surfaces they certainly contain the intersection curve that we are searching
for. The nice feature of the combination (\ref{lower}) is that the maximal degree 
of the polynomials has now decreased to
three. By repeating this procedure using now in the right hand side of Eq.(\ref{lower})
the new polynomials $\tilde{S}_1(\bar{b})$ and $\tilde{S}_2(\bar{b})$  we are able 
to lower the degree in $\bar{b}$ once again by one. After three such steps we
reach a couple of relations that are both linear in $\bar{b}$ and this variable
can finally be eliminated. By extracting the weight $\bar{b}$ from one of the
relations and substituting the result back into the other we obtain 
a polynomial in the variables $a$, $b$ and $c$ that factorizes in terms
of three curves with degrees six, eight and nine. The components with
degree six and nine play the role of extraneous factors and they should
be discarded since our previous rigorous argument tell us that
the sought plane curve should have degree eight. 
The explicit expression for this algebraic plane octic curve is,   
\begin{eqnarray}
\label{curva8}
\mathrm{C}_1(a,b,c) &=& (\varepsilon_1 \varepsilon_2-1)[a^4 + a^2b^2 + b^4]^2+  
(2 + \varepsilon_1^2 - \varepsilon_1 \varepsilon_2 + \varepsilon_2^2)\left [ \varepsilon_2(a^4+a^2b^2+b^4)+abc^2 \right ] abc^2 \nonumber \\
&-&\left [ (\varepsilon_1 \varepsilon_2-2)a^4+(2-\varepsilon_1 \varepsilon_2+\varepsilon_2^2)b^4
-\varepsilon_2^2a^2b^2+2\varepsilon_2abc^2+c^4 \right] c^4,
\end{eqnarray}
and after a systematic use of the constrain (\ref{curva8}) we are able to simplify the expression
for the weight $\bar{b}$ to obtain,
\EQ
\label{bbarA}
\bar{b}=\frac{\left[ (\varepsilon_2-\varepsilon_1)a^2+\varepsilon_2b^2\right]bc^2-a[a^4+a^2b^2+b^4-c^4]}
{\varepsilon_2(a^4+a^2b^2+b^4)+2abc^2}.
\EN

As possible check of the above results we can substitute  
the weight (\ref{bbarA}) in the original
polynomial equations defining the surfaces $S_1$ and $S_2$.
With the help of a symbolic algebra
system one easily find that  
the relations (\ref{surfS1},\ref{surfS2}) become indeed 
proportional to the octic plane curve (\ref{curva8}).

We now turn our attention to discuss the geometric properties of the degree eight
plane curve (\ref{curva8}) governing the integrability of the main branch manifold. 
The basic invariant characterizing a projective algebraic plane curve is 
the topological genus of the respective compact Riemann surface normalization.
In order to compute the genus we need first to identify the singular points
and afterwards to investigate their morphology which include the understanding of possible
infinitesimal neighboring singularities. By solving
the polynomial equations (\ref{sing}) for the plane curve (\ref{curva8}) 
we find a total number of twelve singular points. The first eight of them
are located in the affine plane $c=1$ and they can be expressed as,
\EQ
\label{singaf}
{\mathrm{P}}_{A}= [a_s:b_s:1], 
\EN 
where the values $a_s$ and $b_s$ are the non-null solutions of the following
relations,
\begin{eqnarray}
\label{singaf1}
&& \varepsilon_2^2 a_s^3 + (\varepsilon_1 \varepsilon_2-2)a_sb_s^2 + \varepsilon_2 b_s=0, \\
\label{singaf2}
&& \varepsilon_2^2 b_s^3 +
(2 - \varepsilon_1 \varepsilon_2 + \varepsilon_2^2)a_s^2b_s  + \varepsilon_2 a_s=0.
\end{eqnarray}

The above coupled non-linear equations can be resolved in terms of the roots
of a single univariate polynomial with degree eight. The final answer is somewhat
cumbersome but for sake of completeness it has been presented in Appendix C. 
At this point we recall that a relevant index characterizing 
the morphology a singular point    
is the multiplicity of the singularity. A singular point $[a_0:b_0:c_0]$
is said to have multiplicity $m$ on the plane curve $\mathrm{C}_1(a,b,c)$ if every
partial derivative with respect to the coordinates $a$, $b$ and $c$ up to the order
$m-1$ vanishes at $[a_0:b_0:c_0]$  while at least one 
of order $m$ is non null at this point.
In addition, the singular point is
called ordinary when the factorization of the leading 
term of the expansion of $\mathrm{C}_1(a,b,c)$
around $[a_0:b_0:c_0]$ gives rise to $m$ different terms and thus we dont have
multiple tangents through this point. It turns out that
the Taylor expansion of $\mathrm{C}_1(a,b,1)$ near 
to the affine singularities
defined by Eqs.(\ref{singaf}-\ref{singaf2}) always produces two distinct 
tangents for generic values of the parameters $\varepsilon_1$ and 
$\varepsilon_2$. This means that all the affine singularities are in fact 
ordinary double singular points.   

The remaining four singular points are located at the infinity line $c=0$
being just the zeros of the monomial $\mathrm{C}_1(a,b,0)$. As a consequence they
are independent of $\varepsilon_1$ and $\varepsilon_2$ and their explicit coordinates are,
\EQ
\label{singif}
\mathrm{P}_{\infty}=[\pm 
\exp(\mathrm{i} \frac{\pi}{3}):1:0]~~\mathrm{and}~~ 
[\pm \exp(\mathrm{i} \frac{2\pi}{3}):1:0]. 
\EN

The singularities at infinity behave as double point with
only one tangent but having two branches. They are not 
ordinary singular points and are usually known 
in the literature as tacnodes. Their presence means the existence
of extra neighboring singular points and the plane curve
$\mathrm{C}_1(a,b,c)$ desingularizes only after the
implementation of a sequence of two global birational
transformations often named blowing-ups. The main idea
of this method is to replace a point in the plane by a projective
line which opens more room for the curve to become
non-singular in a higher dimensional space.
For the technical details concerning this approach
in the context of desingularization of an algebraic plane curve
we refer to an overview  by Abhyankar \cite{ABY}. As a
result the resolution
of the singularities of the octic plane 
curve (\ref{curva8}) can be
represented by the diagram,
\EQ
\label{blow}
\renewcommand{\arraystretch}{1.5}
\begin{array}{ccccc}
{{\bf{{C}}}}_1 & \overset{\pi_2}{\longrightarrow} & \tilde{\mathrm{C}}_{1} 
& \overset{\pi_1}{\longrightarrow} & \mathrm{C}_1(a,b,c) .  
\end{array}
\EN
where $\pi_1$ and $\pi_2$ represent two consecutive blowing-ups, the intermediate
curve $\tilde{\mathrm{C}}_1$ carries infinitely near singularities and the
smooth normalization is denoted by
${{\bf{{C}}}}_1$. 

Under the transformation $\pi_1$ all the affine singular points are mapped
onto simple non-singular points while the  
singularities at the infinity line become four ordinary double points. 
The latter are called infinitesimal 
neighboring singular points now sited in
$\tilde{\mathrm{C}}_1$ which are finally resolved 
with the help of the second map $\pi_2$. The genus of the
curve normalization can be computed using the following standard 
formula valid for plane curves,
\EQ
\label{genus}
g(\mathrm{\bf{C}_1})= \frac{(D-1)(D-2)}{2} - \sum_{P} \frac{m_P (m_P-1)}{2},
\EN
where $D$ is the degree of the curve and $m_P$ denotes the multiplicity
of the singular point $P$. The sum is taken over all the singular points
on the curves that are infinitesimal neighbors of $\mathrm{C}_1(a,b,c)$. 

At this point we know that all the singular points have multiplicity $m_P=2$
including those in the infinitesimal neighborhood 
of the singular points at the infinity line. Considering this information 
in Eq.(\ref{genus}) we obtain that the genus is, 
\EQ	
g(\mathrm{\bf{C}_1})= \frac{7 \times 6}{2}   -\underbrace{12 \times 1}_{\pi_1}- \underbrace{4 \times 1 }_{\pi_2} =5,
\EN
whose value can be confirmed within symbolic algebra packages capable
of computing the genus of plane curves such as Singular \cite{SIN}.

In general, canonical genus five curves are known to be realized as the complete 
intersection of three quadrics in $\mathbb{CP}^4$ but they also can degenerate
in either hyperelliptic or trigonal curves, see for example \cite{RIC}. 
One effective way to shed some light on 
the actual class of the algebraic curve (\ref{curva8}) is 
to investigate possible mappings to other curves with lower genus. Exploring
the fact that the polynomial (\ref{curva8}) depends only on even powers of
the weight $c$ we are able to establish the following regular map,  
\EQ
\label{map}
\renewcommand{\arraystretch}{1.5}
\begin{array}{ccc}
\mathrm{C}_1(a,b,c) \subset \mathbb{CP}^2 &~~~ \overset{\phi}{\longrightarrow}~~~ & \mathrm{Q}_1(x,y,z) \subset \mathbb{CP}^2 \\
\left[a:b:c\right] & \longmapsto & \left[a^2:ab:c^2\right],
\end{array}
\EN
where the image of the map $\phi$ is the algebraic curve $\mathrm{Q}_1(x,y,z)$ defined by,
\begin{eqnarray}
Q_1(x,y,z) &=&(\varepsilon_1\varepsilon_2-1)[x^4+x^2y^2+y^4]^2 
+(2+\varepsilon_1^2-\varepsilon_1\varepsilon_2+\varepsilon_2^2)[\varepsilon_2(x^4+x^2y^2+y^4)+x^2yz]x^2yz \nonumber \\
&-&\left[ (\varepsilon_1\varepsilon_2-2)x^4+(2-\varepsilon_1\varepsilon_2+\varepsilon_2^2)y^4-
\varepsilon_2^2x^2y^2+2\varepsilon_2x^2yz+x^2z^2 \right]x^2z^2. 
\end{eqnarray}

The degree of the above map is two because this is
the cardinality of the fiber $\phi^{-1}(\mathrm{P})$ for 
a generic point $\mathrm{P} \in \mathrm{Q}_1(x,y,z)$. To make further progress on 
this admissible double cover we need to compute the genus of the target
curve. In this sense we find that $\mathrm{Q}_1(x,y,z)$ has eight 
ordinary singular points and one singularity resembling 
the tacnode behaviour but with
higher multiplicity $m_P=4$. 
For sake of completeness the technical
details entering this analysis have been summarized in Appendix C. The 
curve $\mathrm{Q}_1(x,y,z)$  
desingularizes once again after a sequence of 
two blowing-ups and the desingularization
diagram is similar to that shown in (\ref{blow}). Denoting
the corresponding blowing-ups by $\bar{\pi}_1$ and $\bar{\pi}_2$
we find that genus of this curve is given by,
\EQ	
\label{genus1}
g(\mathrm{\bf{Q}_1})= \frac{7 \times 6}{2} -\underbrace{(8 \times 1+ 1 \times 6) }_{\bar{\pi}_1}- \underbrace{1 \times 6 }_{\bar{\pi}_2} =1,
\EN
where $\mathrm{\bf{Q}}_1$ denotes the normalization of $\mathrm{Q}_1(x,y,z)$ which turns out to be an elliptic curve.

Now, putting all these
information together we can establish
the following commutative diagram,
\EQ
\label{diagrama}
\renewcommand{\arraystretch}{1.5}
\begin{array}[c]{ccc}
{\bf{C}}_1~~~~~ & \stackrel{\psi}{\longrightarrow} & {\bf{Q}}_1\\
\Big\downarrow\scriptstyle{\pi_1 \circ \pi_2} &&~~~~ \Big\downarrow\scriptstyle{\bar{\pi}_1 \circ \bar{\pi}_2}\\
\mathrm{C}_1(a,b,c) & \stackrel{\phi}{\longrightarrow} & \mathrm{Q}_1(x,y,z) .\\
\end{array} 
\EN
where $\psi$ denotes the morphism induced by the map $\phi$.

From the diagram (\ref{diagrama}) we see that the morphism $\psi$ 
has also degree two and therefore
the genus five 
curve $\mathrm{\bf{C}_1}$ admits a double covering
in terms of the smooth elliptic curve $\mathrm{\bf{Q}_1}$. The existence of such
degree two map is known to rule out the possibility that
the octic plane curve (\ref{curva8}) be sited in the space of either hyperelliptic
or trigonal curves \cite{ACOLA}. We then conclude 
that we are in fact dealing with a bielliptic genus five 
curve whose canonical model is 
the complete intersection of three 
independent quadrics in $\mathbb{CP}^4$.

We close this subsection emphasizing that the above discussion 
on the geometric properties of the octic plane curve (\ref{curva8})
is valid for generic points on the two-dimensional 
space generated by the free parameters $\varepsilon_1$
and $\varepsilon_2$. Such family of curves may
degenerated to plane curves with lower genus once we restrict
$\varepsilon_1$ and $\varepsilon_2$ 
to lie on certain specific submanifolds of 
the parameter space. 
This fact can
naturally occur when one of the surfaces (\ref{surfS1},\ref{surfS2})
becomes reducible and as result their intersection will give rise
to reducible plane curves whose components should have degree 
lower than eight. In Appendix B we have in fact remarked that the 
the surface (\ref{surfS2})
can degenerate into the product of two cones over plane conics. This
happens when the parameters $\varepsilon_1$ and $\varepsilon_2$ 
are constrained to sit in the following one-dimensional submanifolds,
\begin{eqnarray}
\label{mani1}
&& \varepsilon_2 \exp(\pm \mathrm{i} \frac{\pi}{6} )
-\varepsilon_1 \exp(\mp \mathrm{i} \frac{\pi}{6} ) \mp 2 \mathrm{i}=0 \\
\label{mani2}
&& \varepsilon_2 \exp(\pm \mathrm{i} \frac{\pi}{6} )
-\varepsilon_1 \exp(\mp \mathrm{i} \frac{\pi}{6} ) \pm 2 \mathrm{i}=0 \\
\label{mani3}
&& 4\varepsilon_1^2 + \varepsilon_1^4 - 2\varepsilon_1^3\varepsilon_2 + 4\varepsilon_2^2 + 3\varepsilon_1^2\varepsilon_2^2 - 
       2\varepsilon_1\varepsilon_2^3 + \varepsilon_2^4 =0.
\end{eqnarray}

It is not difficult to repeat the previous reasoning 
on the intersection of the 
surfaces (\ref{surfS1},\ref{surfS2}) now for the 
particular submanifolds (\ref{mani1}-\ref{mani3}). Due the
factorization of the degree four surface (\ref{surfS2}) one expects
that the result of the intersection will lead us to the 
product of two quartic algebraic
plane curves. This analysis is somehow direct for
the first two manifolds (\ref{mani1},\ref{mani2})
since one of the parameter can be 
linearly eliminated but it is more involved for the
third non-linear submanifold (\ref{mani3}). The technical details 
entering this analysis are presented in Appendix D 
and here we only state our main conclusions. From 
the explicit expressions of these quartic plane curves we conclude
that they are singular.
The singularities associated 
to the linear submanifolds (\ref{mani1},\ref{mani2}) are single tacnodes 
at the infinity line while those related to the third submanifold (\ref{mani3}) are 
constituted of 
two ordinary double points in the affine plane. This means that when the
parameters $\varepsilon_1$ and $\varepsilon_2$ are
constrained in any of the special submanifolds (\ref{mani1}-\ref{mani3})
the main branch integrable vertex model is therefore governed 
by plane quartic curves of genus one.

\subsection{The Special Branch}

The situation for the special branch is much simpler since the weight
$\bar{b}$  can be linearly eliminated from the divisor (\ref{DIV3}). 
Considering the results of Table (\ref{TAB1}) one would think that 
this branch has to be splited in four different 
integrable manifolds associated to the four possible values of the parameter
$\tilde{\Lambda}_2 =\pm \exp(\pm \mathrm{i} \frac{\pi}{6})$.
However, these apparent distinct vertex
models are related to each other and as result we have only one independent
integrable manifold. The multiplicative signs are easily gauged away with the
help of standard gauge transformations. In addition, the vertex models
with $\tilde{\Lambda}_2 =\exp(\mathrm{i} \frac{\pi}{6})$ and  
$\tilde{\Lambda}_2 =\exp(-\mathrm{i} \frac{\pi}{6})$ can then be 
connected by applying the Weyl basis transformation 
$ 1 \leftrightarrow 3$ on the Lax operator (\ref{LAX}). 
Taking this observation into account we find that the 
expression of the weight $\bar{b}$ 
for such single vertex model is,
\EQ
\label{bbarB}
\bar{b} = \frac{\Lambda_0 b c^2}{ a^2 +\Lambda_0b^2}, 
\EN
where we recall that $\Lambda_0=\exp(\pm \mathrm{i} \frac{\pi}{3})$.

In order to obtain the constrain among 
the variables $a$, $b$ and $c$ we substitute the
above expression for the weight 
$\bar{b}$ in the divisor (\ref{DIV4}). After some 
simplifications  
using the identities satisfied by the parameter $\Lambda_0$ we
find the respective algebraic plane curve is given by,
\EQ
\label{curva6}
\mathrm{C}_2(a,b,c)=[a^2+\Lambda_0 b^2](a^4+a^2b^2+\Lambda_4abc^2)+(b^2-a^2)c^4,
\EN
where $\Lambda_4$ is the free parameter.

The sextic plane curve (\ref{curva6}) has three singular 
points being one of them
on the affine plane while the others are sitting 
on the line at infinity. Their explicit coordinates are given by,
\EQ
\mathrm{P}_{A}=[0:0:1],~~\mathrm{P}_{\infty}=[\frac{1}{\Lambda_0}:1:0]~\mathrm{and}~
[-\frac{1}{\Lambda_0}:1:0].
\EN

The singularity at the origin of the affine plane is an ordinary double 
point while the remaining ones behave as tacnodes and 
the sextic curve is desingularized again by a sequence of two
blowing-ups. We can compute the genus of the normalization
$\mathrm{\bf{C}}_2$ of the sextic curve $\mathrm{C}_2(a,b,c)$
along the same lines presented in previous subsection. 
By applying the formula (\ref{genus}) we obtain that
the genus of $\mathrm{\bf{C}}_2$ is,
\EQ	
g(\mathrm{\bf{C}_2})= \frac{5 \times 4}{2}   
-\underbrace{3 \times 1}_{\pi_1}- \underbrace{2 \times 1 }_{\pi_2} =5.
\EN

In what follows we shall argue that the sextic plane curve 
turns out to be a bielliptic genus five curve as well. The first 
step is to note
that the same two sheeted cover we have discussed before can also
be established for the sextic curve (\ref{curva6}), namely 
\EQ
\label{map1}
\renewcommand{\arraystretch}{1.5}
\begin{array}{ccc}
\mathrm{C}_2(a,b,c) \subset \mathbb{CP}^2 &~~~ \overset{\phi}{\longrightarrow}~~~ & \mathrm{Q}_2(x,y,z) \subset \mathbb{CP}^2 \\
\left[a:b:c\right] & \longmapsto & \left[a^2:ab:c^2\right],
\end{array}
\EN
where the expression of the algebraic target curve $\mathrm{Q}_2(x,y,z)$ is, 
\EQ
\label{tarQ2}
\mathrm{Q}_2(x,y,z) =[x^2 + \Lambda_0 y^2](x^4 + x^2y^2 + y^4 +\Lambda_4x^2yz)+(y^2-x^2)x^2z^2. 
\EN

We next observe that the image curve (\ref{tarQ2}) is quadratic in the variable $z$ and
therefore the linear term on this variable can be eliminated by quadrature. In an 
analogy to what has been explained in Appendix B this define a birational transformation
now in the projective space. More precisely, we are able to put forward the following
second mapping,
\EQ
\label{map2}
\renewcommand{\arraystretch}{1.5}
\begin{array}{ccc}
\mathrm{Q}_2(x,y,z) \subset \mathbb{CP}^2 &~~~ \overset{\tilde{\phi}}{\longrightarrow}~~~ 
& \mathrm{\widetilde{Q}}_2(x_1,y_1,z_1) \subset \mathbb{CP}^2 \\
\left[x:y:z\right] & \longmapsto & \left[\tilde{\phi}_1(x,y,z):\tilde{\phi}_2(x,y,z):\tilde{\phi}_3(x,y,z)\right],
\end{array}
\EN
where the polynomial components of the map (\ref{map2}) are given by,
\begin{eqnarray}
\tilde{\phi}_1(x,y,z) &=& x(x^2 + \Lambda_0y^2), \nonumber \\
\label{map3}
\tilde{\phi}_2(x,y,z) &=& y(x^2 + \Lambda_0y^2), \\
\tilde{\phi}_3(x,y,z) &=& -\mathrm{i}\left[(\Lambda_4y(x^2 + \Lambda_0y^2) + 2(y^2 - x^2)z\right], \nonumber
\end{eqnarray}
while the expression of the image curve
${\widetilde{Q}}_2(x_1,y_1,z_1)$ is,
\EQ
\label{curvaQ2ELI}
\mathrm{\widetilde{Q}}_2(x_1,y_1,z_1)=x_1^2z_1^2 - \frac{4}{\Lambda_0}y_1^4 - (4\Lambda_0 - \Lambda_4^2)x_1^2y_1^2 +4x_1^4
\EN

Direct inspection of the quartic plane curve 
(\ref{curvaQ2ELI}) reveals us that 
it can readily be brought 
into the Jacobi's form of an elliptic curve.
Considering the combination of the above two maps
we are able to build up the following diagram,
\EQ
\label{diagrama1}
\renewcommand{\arraystretch}{1.5}
\begin{array}[c]{ccc}
{\mathrm{C}}_2(a,b,c) & \stackrel{\phi}{\longrightarrow} & {\mathrm{Q}}_2(x,y,z)\\
&\searrow &\Big\downarrow\scriptstyle{\tilde{\phi}}\\
&&\mathrm{\widetilde{Q}}_2(x_1,y_1,z_1) 
\end{array} 
\EN

The degree of the composition of transformations among varieties 
having same dimension is the product 
of the degrees of the individual mappings and from this fact it follows
that $\mathrm{deg}(\tilde{\phi} \circ \phi)=2$. This means that the 
diagram (\ref{diagrama1}) represents
a direct double cover mapping from the singular genus five curve
${\mathrm{C}}_2(a,b,c)$ to a quartic singular curve with genus one.
As before this shows that the sextic plane curve (\ref{curva6}) sits
in the moduli space of bielliptic curves.

Finally, we remark that another direct consequence of the diagram
(\ref{diagrama1}) is that at the special values 
$\Lambda_4^2=12 \Lambda_0, -4 \Lambda_0$ the sextic curve (\ref{curva6})
should degenerate into a plane curve of genus one.
This is because at these parameter values the discriminant of the elliptic
curve $\mathrm{\widetilde{Q}}_2(x_1,y_1,z_1)$ is zero and as result
we have a double cover mapping
$\tilde{\phi} \circ \phi$ from the original plane sextic curve  
to a rational conic curve.
In fact, for $\Lambda_4^2=12 \Lambda_0$ and $-4 \Lambda_0$ 
we verified that sextic curve $\mathrm{C}_2(a,b,c)$ factorizes in terms
of the product of two non-singular cubic plane curves.

\section{The Integrable Lattice Models}
\label{LAXRH}
We now gathered the basic ingredients to present our main results
in terms of the language often used in the modern algebraic
theory of exactly solvable lattice systems. At this point
we know that the Boltzmann weights of the uncovered vertex
models sit on algebraic plane curves and in principle
they can be uniformized by means of a single spectral parameter.
For future convenience we rewrite the Yang-Baxter algebra
in the following more general form,
\EQ
\label{YBN}
\mathrm{R}_{\alpha,\beta}(\lambda,\mu) \mathrm{L}_{\alpha,k}(\lambda) \mathrm{L}_{\beta,k}(\mu) =
\mathrm{L}_{\beta,k}(\mu) \mathrm{L}_{\alpha,k}(\lambda) \mathrm{R}_{\alpha,\beta}(\lambda,\mu),~ k=1,\cdots,N,
\EN
where $\alpha, \beta$ indicate the $\mathrm{R}$-matrix action on the auxiliary spaces while the index $k$ represents
the quantum spaces. The spectral parameters are denoted by $\lambda$ and $\mu$.

We start by presenting the explicit expressions for the
Lax operators $\mathrm{L}_{\alpha,k}(\lambda)$ and the $\mathrm{R}$-matrix
$\mathrm{R}_{\alpha,\beta}(\lambda,\mu)$.

\subsection{Lax operator and R-matrix}

Without loss of generality we shall normalize the Lax operator by the
weight $c$. It turns out that for both branches the Lax operator (\ref{LAX})
can be written as follows,
\begin{eqnarray}
\mathrm{L}_{\alpha,k}(\lambda)&=& a(\lambda)[e_{11}^{(\alpha)} \otimes e_{11}^{(k)} +e_{33}^{(\alpha)} \otimes e_{33}^{(k)}]
+b(\lambda)[e_{11}^{(\alpha)} \otimes e_{22}^{(k)} +e_{33}^{(\alpha)} \otimes e_{22}^{(k)}]+
\bar{b}(\lambda)[e_{22}^{(\alpha)} \otimes e_{11}^{(k)} +e_{22}^{(\alpha)} \otimes e_{33}^{(k)}] \nonumber \\
&+&[e_{12}^{(\alpha)} \otimes e_{21}^{(k)} +e_{21}^{(\alpha)} \otimes e_{12}^{(k)} 
+e_{23}^{(\alpha)} \otimes e_{32}^{(k)} +e_{32}^{(\alpha)} \otimes e_{23}^{(k)}] 
+d(\lambda)[e_{12}^{(\alpha)} \otimes e_{32}^{(k)} +e_{23}^{(\alpha)} \otimes e_{21}^{(k)}] \nonumber \\
&+&\exp(\pm \mathrm{i} \frac{\pi}{3}) d(\lambda)[e_{21}^{(\alpha)} \otimes e_{23}^{(k)} +e_{32}^{(\alpha)} \otimes e_{12}^{(k)}]
+ f(\lambda)[e_{11}^{(\alpha)} \otimes e_{33}^{(k)} +e_{33}^{(\alpha)} \otimes e_{11}^{(k)}]
+g(\lambda)[e_{22}^{(\alpha)} \otimes e_{22}^{(k)}] \nonumber \\
&+&[a(\lambda)+\exp(\mp \mathrm{i} \frac{\pi}{3})f(\lambda)]
[e_{13}^{(\alpha)} \otimes e_{31}^{(k)}]+
[a(\lambda)+\exp(\pm \mathrm{i} \frac{\pi}{3})f(\lambda)][e_{31}^{(\alpha)} \otimes e_{13}^{(k)}]
\end{eqnarray}

The structure of the weights $\bar{b}(\lambda)$, $d(\lambda)$, $f(\lambda)$ 
and $g(\lambda)$ are however branch dependent as well as the underlying algebraic plane curves.
For the main branch the weight $\bar{b}(\lambda)$ has been already determined by Eq.(\ref{bbarA})   
while the remaining weights are obtained by solving 
the divisors (\ref{INV1},\ref{INV2},\ref{INV5}). Considering
the constrains among the parameters listed in Table (\ref{TAB1}) we find that the final results 
for these weights are:

$\bullet$ The main branch
\begin{eqnarray}
\bar{b}(\lambda) &=&\frac{\left[ (\varepsilon_2-\varepsilon_1)a^2(\lambda)+\varepsilon_2b^2(\lambda)\right]b(\lambda)-
[a^4(\lambda)+a^2(\lambda)b^2(\lambda)+b^4(\lambda)-1]a(\lambda)}
{\varepsilon_2[a^4(\lambda)+a^2(\lambda)b^2(\lambda)+b^4(\lambda)]+2a(\lambda)b(\lambda)}, \\ \nonumber \\
d(\lambda) &=& \pm \sqrt{\frac{\varepsilon_1 \varepsilon_2-1}{\exp(\pm \mathrm{i} \frac{\pi}{3})}}
\frac{b(\lambda)}{[\varepsilon_1 a(\lambda) +\bar{b}(\lambda)]},~~~
f(\lambda) = \frac{b(\lambda) [ \varepsilon_1a(\lambda) b(\lambda) + b(\lambda) \bar{b}(\lambda)-1]}{a(\lambda)[
\varepsilon_1 a(\lambda) +\bar{b}(\lambda)]}, \\ \nonumber \\
\label{gmain}
g(\lambda) &=& \left[ \frac{\varepsilon_1+\varepsilon_2}{\varepsilon_1 \varepsilon_2} \right] b(\lambda)+
\left [ \frac{(\varepsilon_1 \varepsilon_2-1)b(\lambda)}{\varepsilon_1[\varepsilon_1 a(\lambda)+\bar{b}(\lambda)]}
+\frac{\varepsilon_1 a(\lambda)+\bar{b}(\lambda)}{\varepsilon_2 b(\lambda)} \right ]\bar{b}(\lambda),
\end{eqnarray}
where the variables $a(\lambda)$ and $b(\lambda)$ fulfill  
the affine version of the genus five octic plane curve (\ref{curva8}), 
\begin{eqnarray}
\label{curva8a}
\mathrm{C}_2(\lambda) &=& (\varepsilon_1 \varepsilon_2-1)[a^4(\lambda) + a^2(\lambda)b^2(\lambda) + b^4(\lambda)]^2  
-(\varepsilon_1 \varepsilon_2-2)a^4(\lambda)+(2-\varepsilon_1 \varepsilon_2+\varepsilon_2^2)b^4(\lambda) \nonumber \\
&+& (2 + \varepsilon_1^2 - \varepsilon_1 \varepsilon_2 + \varepsilon_2^2)
\left [ \varepsilon_2 \{ a^4(\lambda)+a^2(\lambda)b^2(\lambda)
+b^4(\lambda) \} 
+ a(\lambda)b(\lambda) \right ] a(\lambda)b(\lambda) \nonumber \\ 
&-&\varepsilon_2^2a^2(\lambda)b^2(\lambda)+2\varepsilon_2a(\lambda)b(\lambda)+1. 
\end{eqnarray}

Similar results for the special branch are now obtained 
considering the divisors (\ref{DIV1},\ref{DIV2},\ref{INV5},\ref{bbarB}). They can be solved
linearly and the expressions for the weights are:

$\bullet$ The special branch
\begin{eqnarray}
\bar{b}(\lambda) &=& \frac{b(\lambda)}{\exp(\mp \mathrm{i} \frac{\pi}{3}) a^2(\lambda) +b^2(\lambda)},~~~ 
d(\lambda)= \frac{b(\lambda)}{a(\lambda)}, \\ \nonumber \\
g(\lambda) &=& \frac{a^2(\lambda)+\exp(\pm \mathrm{i} \frac{2\pi}{3}) b^2(\lambda)} 
{a(\lambda)[a^2(\lambda)+\exp(\pm \mathrm{i} \frac{\pi}{3}) b^2(\lambda)]},~~~
f(\lambda) = \frac{b^2(\lambda)}{a(\lambda)}, 
\end{eqnarray}
where the variables $a(\lambda)$ and $b(\lambda)$ satisfy
the affinization  of the genus five sextic plane curve (\ref{curva6}), 
\EQ
\label{curva6a}
\mathrm{C}_2(\lambda)=[a^2(\lambda)+\exp(\pm \mathrm{i} \frac{\pi}{3}) b^2(\lambda)][a^4(\lambda)+a^2(\lambda)b^2(\lambda)
+\Lambda_4a(\lambda)b(\lambda)]+b^2(\lambda)-a^2(\lambda)
\EN

An important property of the above Lax operators is that there exists a special value
for the spectral parameter $\lambda$ in which 
they become proportional to the permutator. This turns out to be the point $\lambda_0$
on both curves (\ref{curva8a},\ref{curva6a}) in which $a(\lambda_0)=1$ and $b(\lambda_0)=0$.
In fact, considering
the expressions for the weights $\bar{b}(\lambda)$, $d(\lambda)$, $f(\lambda)$
and $g(\lambda)$ 
at the value $\lambda_0$ we obtain\footnote{Note that for the main branch the direct limit
$\displaystyle{\lim_{\lambda \rightarrow \lambda_0}} g(\lambda)$ is indefinite. This indeterminacy is
evaluated by substituting the expression for $\bar{b}(\lambda)$ in Eq.(\ref{gmain}) 
and afterwards the high weights powers are reduced with  
the help of the curve constrain (\ref{curva8a}).
After carrying on these simplifications we find indeed that $g(\lambda_0)=1$},   
\EQ
\mathrm{L}_{\alpha,k}(\lambda_0)= \mathcal{P}_{\alpha,k} =\sum_{i,j=1}^{3} e_{ij}^{(\alpha)} \otimes e_{ji}^{(k)}
\EN

Let us now turn our attention to the $\mathrm{R}$-matrix. After some cumbersome
simplifications we find that the 
$\mathrm{R}$-matrix has a universal 
structure for both branches
once we write its elements in terms of an enlarged set 
of weights given by $a(\lambda)$, $b(\lambda)$, $\bar{b}(\lambda)$, 
$d(\lambda)$ and $f(\lambda)$. The form
of the $\mathrm{R}$-matrix becomes similar to that of Lax operators, namely
\begin{eqnarray}
\mathrm{R}_{\alpha,\beta}(\lambda,\mu)&=& {\bf{a}}(\lambda,\mu)[e_{11}^{(\alpha)} \otimes e_{11}^{(\beta)} +e_{33}^{(\alpha)} \otimes e_{33}^{(\beta)}]
+{\bf{b}}(\lambda,\mu)[e_{11}^{(\alpha)} \otimes e_{22}^{(\beta)} +e_{33}^{(\alpha)} \otimes e_{22}^{(\beta)}] \nonumber \\
&+& {\bf{\overline{b}}}(\lambda,\mu)[e_{22}^{(\alpha)} \otimes e_{11}^{(\beta)} +e_{22}^{(\alpha)} \otimes e_{33}^{(\beta)}]
+[e_{12}^{(\alpha)} \otimes e_{21}^{(\beta)} +e_{21}^{(\alpha)} \otimes e_{12}^{(\beta)} 
+e_{23}^{(\alpha)} \otimes e_{32}^{(\beta)} +e_{32}^{(\alpha)} \otimes e_{23}^{(\beta)}] \nonumber \\
&+&{\bf{d}}(\lambda,\mu)[e_{12}^{(\alpha)} \otimes e_{32}^{(\beta)} +e_{23}^{(\alpha)} \otimes e_{21}^{(\beta)}] 
+\exp(\pm \mathrm{i} \frac{\pi}{3}) {\bf{d}}(\lambda,\mu)[e_{21}^{(\alpha)} \otimes e_{23}^{(\beta)} +e_{32}^{(\alpha)} \otimes e_{12}^{(\beta)}] \nonumber \\
&+& {\bf{f}}(\lambda,\mu)[e_{11}^{(\alpha)} \otimes e_{33}^{(\beta)} +e_{33}^{(\alpha)} \otimes e_{11}^{(\beta)}]
+{\bf{g}}(\lambda,\mu)[e_{22}^{(\alpha)} \otimes e_{22}^{(\beta)}] \nonumber \\
&+&[{\bf{a}}(\lambda,\mu)+\exp(\mp \mathrm{i} \frac{\pi}{3}){\bf{f}}(\lambda,\mu)]
[e_{13}^{(\alpha)} \otimes e_{31}^{(\beta)}]+
[{\bf{a}}(\lambda,\mu)+\exp(\pm \mathrm{i} \frac{\pi}{3}){\bf{f}}(\lambda,\mu)][e_{31}^{(\alpha)} \otimes e_{13}^{(\beta)}] \nonumber \\
\end{eqnarray}
where the matrix elements 
${\bf a}(\lambda,\mu)$, 
${\bf b}(\lambda,\mu)$, 
${\bf{\overline{b}}}(\lambda,\mu)$ 
${\bf d}(\lambda,\mu)$, 
${\bf f}(\lambda,\mu)$, and
${\bf g}(\lambda,\mu)$ are given by,
\begin{eqnarray}
{\bf a}(\lambda,\mu) &=& \frac{a(\lambda) + [\bar{b}(\lambda)a(\mu)- a(\lambda)\bar{b}(\mu)]b(\mu)}{a(\mu)}, \\ \nonumber \\
{ \bf b}(\lambda,\mu) &=& 
\frac{[1 
- b(\mu)\bar{b}(\mu)]a(\lambda)b(\lambda)-
[1 - b(\lambda)\bar{b}(\lambda)]a(\mu)b(\mu)}{a(\lambda)a(\mu)}, 
\\ \nonumber \\
{\bf{\overline{b}}}(\lambda,\mu) &=& \bar{b}(\lambda)a(\mu) - a(\lambda)\bar{b}(\mu), \\ \nonumber \\
{\bf d}(\lambda,\mu) &=& \frac{d(\lambda)a(\mu)[\bar{b}(\lambda)a(\mu) - a(\lambda)\bar{b}(\mu)]}{f(\lambda)a(\mu)b(\mu) + 
      \bar{b}(\lambda)[1-b(\mu)\bar{b}(\mu)]}, \\ \nonumber \\
{\bf f}(\lambda,\mu) &=& \frac{a(\mu)[\bar{b}(\lambda)a(\mu) - a(\lambda)\bar{b}(\mu)][f(\lambda)a(\mu) - \bar{b}(\lambda)\bar{b}(\mu)]}
{f(\lambda)a(\mu)b(\mu) + 
      \bar{b}(\lambda)[1-b(\mu)\bar{b}(\mu)]}, \\ \nonumber \\
{\bf{g}}(\lambda,\mu) &=& -{\bf{d}}(\lambda,\mu) \frac{[{\bf{f}}(\mu,\lambda)+{\bf{a}}(\mu,\lambda)]}{{\bf{d}}(\mu,\lambda)}. 
\end{eqnarray}

By construction this $\mathrm{R}$-matrix satisfies the unitarity property 
which now can be stated as,
\EQ
\mathrm{R}_{\alpha,\beta}(\lambda,\mu)\mathrm{R}_{\beta,\alpha}(\mu,\lambda)= a(\lambda,\mu) a(\mu,\lambda)  
{\mathrm{I}_3} \otimes {\mathrm{I}_3},
\EN
being also 
a regular matrix at the point $\lambda_0$,
$\mathrm{R}_{\alpha,\beta}(\lambda_0,\lambda_0)={\mathcal{P}}_{\alpha,\beta}$. 

In addition to that, we have 
verified by means of the algebraic procedure
explained in Appendix A that the $\mathrm{R}$-matrix satisfies the famous Yang-Baxter equation, 
\EQ
\label{YBEN1}
\mathrm{R}_{12}(\lambda_1,\lambda_2) \mathrm{R}_{13}(\lambda_1,\lambda_3) \mathrm{R}_{23}(\lambda_2, \lambda_3) =
\mathrm{R}_{23}(\lambda_2,\lambda_3) \mathrm{R}_{13}(\lambda_1,\lambda_3) \mathrm{R}_{12}(\lambda_1, \lambda_2)
\EN
which is sufficient condition for the associativity of the Yang-Baxter algebra (\ref{YBN}).

We finally remark that given a solution of the Yang-Baxter equation we can easily produce
other equivalent multiparametric solutions by means of the so-called twist transformations. 
These are natural symmetries underlying the Yang-Baxter algebra and for a more detailed 
discussion see for example references \cite{DEV}. 
It turns out
that a special type of twist will be helpful to make a correspondence between our respective quantum
spin-$1$ chains and that derived in the work by Alcaraz and Bariev \cite{ALBA}.
We find that in our case the most general diagonal twist that is compatible with integrability
has the following form,
\EQ
\label{twist}
{\cal{G}}(\tau_1,\tau_2,\tau_3)=\mathrm{diag}(1,\tau_1,\tau_1^2|\tau_2,\tau_3,\frac{\tau_3^2}{\tau_2}|
\tau_2^2,\frac{\tau_3^2}{\tau_1},\frac{\tau_3^4}{\tau_1^2\tau_2^2})
\EN
where $\tau_1$, $\tau_2$ and $\tau_3$ are free additional parameters. 

It can be checked easily that the
following transformed Lax operator and $\mathrm{R}$-matrix 
\EQ
{\cal{G}}_{k,\alpha}(\tau_1,\tau_2,\tau_3)\mathrm{L}_{\alpha,k}(\lambda) 
{\cal{G}}_{\alpha,k}^{-1}(\tau_1,\tau_2,\tau_3)~~~\mathrm{and}~~~
{\cal{G}}_{\beta,\alpha}(\tau_1,\tau_2,\tau_3)\mathrm{R}_{\alpha,\beta}(\lambda,\mu) 
{\cal{G}}_{\alpha,\beta}^{-1}(\tau_1,\tau_2,\tau_3)
\EN
is still another solution for
the Yang-Baxter algebra(\ref{YBN}). 
 
\subsection{Spin Chain Hamiltonians}

The expansion of the 
logarithm of the transfer matrix $\mathrm{T}(\lambda)$ around
the regular point $\lambda_0$ is known to produce a set of mutually commuting operators. 
Of particular interest is the Hamiltonian describing the interaction of nearest 
neighbors spins variables on the lattice,
\EQ
\mathrm{H}= \frac{\partial}{\partial \lambda} \ln \mathrm{T}(\lambda)|_{\lambda=\lambda_0}=\sum_{k=1}^{N} \mathrm{H}_{k,k+1} 
\EN
where $\mathrm{H}_{k,k+1}= \mathcal{P}_{k,k+1} \frac{\partial}{\partial \lambda} \mathrm{L}_{k,k+1}(\lambda)|_{\lambda=\lambda_0}$  and
boundary periodic conditions $\mathrm{H}_{N,N+1}=\mathrm{H}_{N,1}$ are assumed.

In order to compute the Hamiltonian we just need to take the derivative on the 
Lax operators at the point $\lambda=\lambda_0$ and impose 
that $a(\lambda_0)=1$ and $b(\lambda_0)=0$.
The derivatives of the weights $a(\lambda)$ and $b(\lambda)$ are then related under derivation
of the algebraic curves (\ref{curva8a},\ref{curva6a}) and evaluating 
the results at $\lambda=\lambda_0$. In what follows we shall list the final expressions
for the two-body Hamiltonians in the Weyl basis: 

$\bullet$ The main branch

The spin chain for the main branch is given by,
\begin{eqnarray}
\label{spinA}
\mathrm{H}_{k,k+1}^{(\pm)}(\varepsilon_1,\varepsilon_2) &=& \mathrm{J}_1[e_{11}^{(k)} e_{11}^{(k+1)} +e_{33}^{(k)} e_{33}^{(k+1)}]
-[e_{21}^{(k)} e_{12}^{(k+1)}+e_{23}^{(k)} e_{32}^{(k+1)}]
-\frac{\varepsilon_2}{\varepsilon_1}[e_{12}^{(k)} e_{21}^{(k+1)}+e_{32}^{(k)} e_{23}^{(k+1)}] \nonumber \\
&+&[\frac{\exp(\pm \mathrm{i} \frac{\pi}{3})}{\varepsilon_1}+\mathrm{J}_1]e_{11}^{(k)} e_{33}^{(k+1)}
+[\frac{\exp(\mp \mathrm{i} \frac{\pi}{3})}{\varepsilon_1}+\mathrm{J}_1]e_{33}^{(k)} e_{11}^{(k+1)} 
+\frac{1}{\varepsilon_1}[e_{13}^{(k)} e_{31}^{(k+1)}+e_{31}^{(k)} e_{13}^{(k+1)}] \nonumber \\
&+& \mathrm{J}_2\exp(\pm \mathrm{i} \frac{\pi}{6})[e_{12}^{(k)} e_{32}^{(k+1)}+e_{21}^{(k)} e_{23}^{(k+1)}]
+\mathrm{J}_2 \exp(\mp \mathrm{i} \frac{\pi}{6})[e_{23}^{(k)} e_{21}^{(k+1)}+e_{32}^{(k)} e_{12}^{(k+1)}] \nonumber \\
&-&[\frac{1}{\varepsilon_1}+\mathrm{J}_1]e_{22}^{(k)} e_{22}^{(k+1)}
\end{eqnarray}
where the dependence of the couplings $\mathrm{J}_1$ and $\mathrm{J_2}$ on 
the free parameters $\varepsilon_1$ and $\varepsilon_2$ are,
\EQ
\mathrm{J}_1= \frac{\varepsilon_1^2 - \varepsilon_1\varepsilon_2 + \varepsilon_2^2}{4\varepsilon_1}~~~\mathrm{and}~~~
\mathrm{J}_2=\frac{\sqrt{\varepsilon_1\varepsilon_2-1}}{\varepsilon_1}.
\EN

We now remark that the one-parameter
integrable spin-$1$ chain found previously by 
Alcaraz and Bariev \cite{ALBA} is in fact a  
particular case of our Hamiltonian (\ref{spinA}) when
the respective parameters sit on the subspace
$\varepsilon_1=\pm \varepsilon_2$.
Let us denote the two-body Hamiltonian derived in the work \cite{ALBA} by 
$\bar{H}_{k,k+1}(t_p,\epsilon)$ where $\epsilon=\pm$ and $t_p$ 
is the single free parameter 
in the notation of this reference \footnote{F.C. Alcaraz informed us 
that the coupling $u$ in the work \cite{ALBA} should be read as 
$u=\frac{\epsilon t_p}{2} +\frac{(2-\epsilon)}{2t_p}$.}.
With help of a special case of the twisted 
transformation (\ref{twist}) we have been able
to verify the following correspondence 
among two-body operators,
\begin{eqnarray}
\label{corres}
-\bar{H}_{k,k+1}(t_p,\epsilon) &=& \frac{
1}{\sqrt{\epsilon}}{\cal{G}}_{k+1,k}(1,\frac{1}{\sqrt{\epsilon}},\epsilon) 
\mathrm{H}_{k,k+1}^{(+)}\left (-\frac{\sqrt{\epsilon}}{t_p},  
-\frac{1}{\sqrt{\epsilon}t_p} \right) 
{\cal{G}}_{k+1,k}^{-1}(1,\frac{1}{\sqrt{\epsilon}},\epsilon) +\frac{(\epsilon -2)}{4 t_p}[S^z_k+S^z_{k+1}] \nonumber \\
&+& \frac{\mathrm{i} \sqrt{3} t_p}{4 \epsilon}[S^z_k-S^z_{k+1}]  
-\frac{\mathrm{i} \sqrt{3} t_p}{4 \epsilon}[(S^z_k)^2-(S^z_{k+1})^2] 
\end{eqnarray}
where $S^z_k=e_{11}^{(k)}-e_{33}^{(k)}$ denotes the azimuthal 
component of the spin-$1$ operator. The second term in (\ref{corres}) is proportional to
the azimuthal magnetic field and can always be added 
since it commutes with Hamiltonian while the last two terms vanish under periodic boundary
condition and do not contribute to the volume Hamiltonian. We note from Eq.(\ref{corres}) 
that the twist was only necessary to fit the case $\epsilon=-1$.

Of course one can use the more general twist 
transformation (\ref{twist}) to generate
a family of exactly solvable multiparametric Hamiltonians. Because this is
a diagonal twist it does not spoil the $\mathrm{U}(1)$ symmetry and the diagonalization
of the respective vertex model transfer matrix  
could in principle be tackled by general algebraic framework 
proposed in \cite{MEMA}.

$\bullet$ The Special branch

In this case we have one-parameter spin chain Hamiltonian and the expression for the
corresponding two-body operator is, 
\begin{eqnarray}
\label{spinB}
\mathrm{H}_{k,k+1}^{(\pm)}(\Lambda_4) & =& \frac{\Lambda_4}{4}[e_{11}^{(k)} e_{11}^{(k+1)} +e_{33}^{(k)} e_{33}^{(k+1)}
+e_{11}^{(k)} e_{33}^{(k+1)} +e_{33}^{(k)} e_{11}^{(k+1)}-e_{22}^{(k)} e_{22}^{(k+1)}] 
-[e_{23}^{(k)} e_{21}^{(k+1)}+e_{32}^{(k)} e_{12}^{(k+1)}] \nonumber \\
&-&[e_{21}^{(k)} e_{12}^{(k+1)}+e_{23}^{(k)} e_{32}^{(k+1)}]
-\exp(\pm \mathrm{i} \frac{\pi}{3} )[e_{12}^{(k)} e_{32}^{(k+1)}+e_{21}^{(k)} 
e_{23}^{(k+1)}+e_{12}^{(k)} e_{21}^{(k+1)}+e_{32}^{(k)} e_{23}^{(k+1)}] \nonumber \\ 
\end{eqnarray}

\section{Conclusions}
\label{conclu}

In this paper we have investigated the Yang-Baxter algebra for three-state vertex model
whose statistical configurations are invariant by the $\mathrm{U}(1)$ invariance but break
in an explicit way the parity-time reversal symmetry. We argued that the assumption 
of unitarity of
the respective $\mathrm{R}$-matrix imposes us that the functional
equations derived from the Yang-Baxter algebra are anti-symmetrical on the exchange
of the Boltzmann weights of distinct Lax operators. This property provides us 
the means to disentangle 
involved high degree functional
relations in a rather systematic way.  
The integrable manifolds are found by
intersecting a number of prime divisors associated to polynomial equations 
which are naturally separable on the distinct weights labels. We have been
able to uncover two families of integrable nineteen vertex models whose
weights are lying on bielliptic algebraic curves of genus five. 
For the family having two free parameters this comes about after 
dealing with the problem of the
intersection of two projective surfaces: one of them a rational
cubic surface and the other a cone generated by an elliptic curve. 
We have pointed out that genus five bielliptic curves can generate 
to standard elliptic curves when the respective free parameters 
are restricted to particular subspaces. 

The Lax operators have a regular point in which they
become proportional to the permutator and the respective
two families of exactly solvable quantum spin-$1$ chains have been
computed. We have found that our two-parameter Hamiltonian family
generalizes the integrable one-parameter spin-$1$ chain discovered by
Alcaraz and Bariev \cite{ALBA}. We exhibit a relationship between
these Hamiltonians when our free parameters are restricted to the
subspace $\varepsilon_1 =\pm \varepsilon_2$. We have found 
that the $\mathrm{R}$-matrix has the same general 
form for both family of vertex models provided we write it as 
function of a suitable subset of Boltzmann weights. The $\mathrm{R}$-matrix
is non-additive with respect to the spectral parameters 
and we have verified that it satisfies the Yang-Baxter equation by
means of computer algebra system.

A natural question to be asked is whether these integrable vertex models
admit an adequate description in the framework of quantum groups such
as turned out to be the case of chiral Potts model \cite{BAZ}. We think
that a possible hint into this direction comes from the degeneration
of the octic plane curve (\ref{curva8}) into a rational curve for specific
values of the parameters $\varepsilon_1$ and $\varepsilon_2$ discussed
in Appendix D. This fact has motivated us to search for a relation 
among the main branch Hamiltonian at such particular parameter values
and known rational quantum spin-$1$ chains. 
To this end we have been able to relate  
the operator ${\mathrm{H}}^{(\pm)}(\varepsilon_1=2,\varepsilon_2=2)$ defined by Eq.(\ref{spinA})
to that of the trigonometric spin-$1$ chain based on the quantum superalgebra
$\mathrm{U}_{\bar{q}}[\mathrm{Osp}(1|2)]$, often referred as 
the Izergin-Korepin model \cite{IK},
when the deformation 
parameter is ${\bar{q}}=\exp(\mp \mathrm{i} \frac{\pi}{3})$. This suggests 
that the vertex models obtained in this paper may be also originated
from some non-generic three-dimensional representation of the  
$\mathrm{U}_{\bar{q}}[\mathrm{Osp}(1|2)]$ superalgebra probably at roots of unity.
The immediate difficult would be to find the appropriate representation that
is able to reproduce the pertinent complete intersection the 
three quadrics by means of the quantum group machinery. 
Hopefully, this observation will prompt further investigations on
other mathematical properties that are hidden in these vertex models. 

Finally, we expect that the approach used in this paper to solve a number of
entangled functional equations could also be applied to study the Yang-Baxter algebra
associated to generalized nineteen vertex models or even high-state vertex models.
We plan to investigate some of these problems in future works.

\section*{Acknowledgments}
This work has been support by the Brazilian Research Councils CNPq and FAPESP.
I owe an enormous debt of gratitude to Daniel Levcovitz  and 
Herivelto Borges for innumerable discussions on many  
algebraic geometry topics helpful to this paper. Special thanks goes to Daniel Levcovitz
for calling my attention to the Singular computer algebra system. 

\addcontentsline{toc}{section}{Appendix A}
\section*{\bf Appendix A: Extra Functional Relations}
\setcounter{equation}{0}
\renewcommand{\theequation}{A.\arabic{equation}}

In this appendix we present the remaining functional relations 
coming from the Yang-Baxter algebra (\ref{YB}) not presented 
in the main text. Besides the relations having four terms
explicitly exhibited in section \ref{secfour} we have the following extra
twelve equations,
\begin{eqnarray}
\label{fourextra1}
&& \mathbf{a} \bar{h}^{'} {a}^{''} -\mathbf{\bar{d}} c^{'} \bar{d}^{''} -\mathbf{f} \bar{h}^{'} f^{''} 
-\mathbf{\bar{h}} {a}^{'} {\bar{h}}^{''}=0, \\
&& \mathbf{h} f^{'} {a}^{''} -\mathbf{d} c^{'} \bar{d}^{''} -\mathbf{h}a^{'} f^{''} -\mathbf{f} {h}^{'} \bar{h}^{''}=0, \\
&& \mathbf{\bar{h}} f^{'} {a}^{''} -\mathbf{\bar{d}} c^{'} d^{''} -\mathbf{\bar{h}}a^{'} f^{''} -\mathbf{f} {\bar{h}}^{'} {h}^{''}=0, \\
&& \mathbf{d} c^{'} \bar{d}^{''} +\mathbf{h} \bar{h}^{'} f^{''} +\mathbf{f}a^{'} \bar{h}^{''} -\mathbf{a} {f}^{'} {\bar{h}}^{''}=0, \\
&& \mathbf{\bar{d}} c^{'} {d}^{''} +\mathbf{\bar{h}} {h}^{'} f^{''} +\mathbf{f}a^{'} {h}^{''} -\mathbf{a} {f}^{'} {h}^{''}=0, \\
&& \mathbf{d} c^{'} {d}^{''} +\mathbf{h} {\bar{h}}^{'} h^{''} 
-\mathbf{\bar{h}} {h}^{'} \bar{h}^{''} 
-\mathbf{\bar{d}} c^{'} {\bar{d}}^{''}=0,\\ 
&& \mathbf{a} d^{'} {c}^{''} -\mathbf{g} c^{'} d^{''} -\mathbf{\bar{d}}h^{'} f^{''} -\mathbf{d} {a}^{'} h^{''}=0, \\
&& \mathbf{a} \bar{d}^{'} {c}^{''} -\mathbf{g} c^{'} \bar{d}^{''} -\mathbf{d} \bar{h}^{'} f^{''} 
-\mathbf{\bar{d}} {a}^{'} \bar{h}^{''}=0, \\
&& \mathbf{c} d^{'} {a}^{''} -\mathbf{h} a^{'} d^{''} -\mathbf{f}h^{'} \bar{d}^{''} -\mathbf{d} {c}^{'} g^{''}=0, \\
&& \mathbf{c} \bar{d}^{'} {a}^{''} -\mathbf{\bar{h}} a^{'} \bar{d}^{''} -\mathbf{f}\bar{h}^{'} d^{''} 
-\mathbf{\bar{d}} {c}^{'} g^{''}=0, \\
&& \mathbf{c} f^{'} {c}^{''} -\mathbf{d} g^{'} {\bar{d}}^{''} -\mathbf{h}c^{'} f^{''} 
-\mathbf{f} {c}^{'} \bar{h}^{''}=0, \\
\label{fourextra2}
&& \mathbf{c} f^{'} {c}^{''} -\mathbf{\bar{d}} g^{'} d^{''} -\mathbf{\bar{h}}c^{'} f^{''} 
-\mathbf{f} {c}^{'} h^{''}=0. 
\end{eqnarray}

In addition to that we have also functional relations involving five terms. Contrary to
what happen so far their total 
number remains unchanged after the solution of the two terms functional relations. Following Table 
(\ref{TAB1}) 
we have twelve functional equations which are given by,
\begin{eqnarray}
\label{fiveextra1}
&& \mathbf{b} c^{'} \bar{b}^{''} +\mathbf{c} g^{'} c^{''} -\mathbf{d} \bar{h}^{'} d^{''} 
-\mathbf{\bar{d}} a^{'} \bar{d}^{''} -\mathbf{g} c^{'} g^{''}=0,\\
&& \mathbf{\bar{d}} \bar{b}^{'} \bar{b}^{''} +\mathbf{\bar{h}} d^{'} c^{''} -\mathbf{g} c^{'} d^{''} 
-\mathbf{\bar{d}} a^{'} f^{''} -\mathbf{d} \bar{h}^{'} h^{''}=0,\\
&& \mathbf{h} \bar{h}^{'} {d}^{''} +\mathbf{f} a^{'} \bar{d}^{''} -\mathbf{b} b^{'} \bar{d}^{''} 
+\mathbf{d} c^{'} g^{''} -\mathbf{c} {d}^{'} \bar{h}^{''}=0,\\
&& \mathbf{d} \bar{b}^{'} \bar{b}^{''} +\mathbf{h} \bar{d}^{'} c^{''} -\mathbf{g} c^{'} \bar{d}^{''} 
-\mathbf{d} a^{'} f^{''} -\mathbf{\bar{d}} {h}^{'} \bar{h}^{''}=0,\\
&& \mathbf{\bar{b}} c^{'} {b}^{''} +\mathbf{c} \bar{h}^{'} c^{''} -\mathbf{\bar{d}} {g}^{'} \bar{d}^{''} 
-\mathbf{f} c^{'} f^{''} -\mathbf{\bar{h}} {c}^{'} \bar{h}^{''}=0,\\
&& \mathbf{\bar{b}} \bar{b}^{'} {\bar{d}}^{''} -\mathbf{g} g^{'} \bar{d}^{''} -\mathbf{d} {c}^{'} {f}^{''} 
+\mathbf{c} \bar{d}^{'} g^{''} -\mathbf{\bar{d}} {c}^{'} \bar{h}^{''}=0,\\
&& \mathbf{\bar{d}} b^{'} b^{''} +\mathbf{g} \bar{d}^{'} c^{''} -\mathbf{f} {c}^{'} {d}^{''} 
-\mathbf{\bar{h}} {c}^{'} \bar{d}^{''} -\mathbf{\bar{d}} {g}^{'} {g}^{''}=0,\\
&& \mathbf{f} a^{'} d^{''} -\mathbf{b} {b}^{'} d^{''} +\mathbf{\bar{h}} {h}^{'} {\bar{d}}^{''} 
+\mathbf{\bar{d}} {c}^{'} {g}^{''} -\mathbf{{c}} {\bar{d}}^{'} {h}^{''}=0,\\
&& \mathbf{b} c^{'} \bar{d}^{''} +\mathbf{c} {g}^{'} c^{''} -\mathbf{d} {a}^{'} {d}^{''} 
-\mathbf{\bar{d}} {h}^{'} {\bar{d}}^{''} -\mathbf{g} {c}^{'} {g}^{''}=0,\\
&& \mathbf{\bar{b}} \bar{b}^{'} d^{''} -\mathbf{g} {g}^{'} d^{''} -\mathbf{\bar{d}} {c}^{'} {f}^{''} 
+\mathbf{c} {d}^{'} {g}^{''} -\mathbf{d} {c}^{'} {h}^{''}=0,\\
&& \mathbf{g} d^{'} c^{''} -\mathbf{h} {c}^{'} d^{''} -\mathbf{f} {c}^{'} {\bar{d}}^{''} 
-\mathbf{d} {b}^{'} {b}^{''} +\mathbf{d} {g}^{'} {g}^{''}=0,\\
\label{fiveextra2}
&& \mathbf{\bar{b}} c^{'} b^{''} +\mathbf{c} {h}^{'} c^{''} -\mathbf{d} {g}^{'} {d}^{''} 
-\mathbf{f} {c}^{'} {f}^{''} -\mathbf{h} {c}^{'} {h}^{''}=0.
\end{eqnarray}

An effective way to check that all the above equations are indeed satisfied for
the main branch is to proceed as follows.  After
extracting linearly the weights $d$, $f$, $g$, $h$ and $\bar{h}$  
from the divisors (\ref{INV1},\ref{INV2},\ref{INV5},\ref{INV6}) the 
functional equations become polynomials only in the variables 
$a$, $b$, $\bar{b}$ and $c$ for both indices labels. In addition 
to that these weights
are constrained  by the remaining divisors (\ref{INV3},\ref{INV4}).
The main idea of our procedure is to replace in a given
functional equation powers of a subset of variables with the help  
of the last two  divisors:

$ \bullet~ \mathrm{Step~ One} $

We have already mentioned that the functional equations depend only on even powers
of the weight $c$. The power $c^2$ can easily be extracted 
from the divisor (\ref{INV3}) or equivalently
from the surface (\ref{surfS1}). 
Denoting this amplitude by $\mathrm{auxc}$ we obtain,
\begin{eqnarray}
\label{auxc}
\mathrm{auxc} &=& \frac{ab^2+\varepsilon_1a^2{\bar{b}}+\varepsilon_2b^2{\bar{b}}+a{\bar{b}}^2}{\varepsilon_2b} 
\end{eqnarray}

Now we inspect the highest power in a given polynomial equation denoted here 
generically by $\mathrm{eq}[*]$. Assuming that this power is for example six the dependence
on the weight $c$ can be systematically replaced using the following Mathematica code,
\begin{eqnarray}
\mathrm{eq}1 &=& \mathrm{Factor}[\mathrm{eq}[*]] \\
\mathrm{eq}2 &=& \mathrm{Factor}[\mathrm{eq}1 ~/.~ \{ [c^{'}]^6 \rightarrow ~\mathrm{auxc}^{'}~ [c^{'}]^4, [c^{''}]^6 \rightarrow \mathrm{auxc}^{''}~ [c^{''}]^4 \}] \\
\mathrm{eq}3 &=& \mathrm{Factor}[\mathrm{eq}2 ~/.~ \{ [c^{'}]^4 \rightarrow \mathrm{auxc}^{'}~ [c^{'}]^2, [c^{''}]^4 \rightarrow \mathrm{auxc}^{''}~ [c^{''}]^2 \}] \\
\mathrm{eq}4 &=& \mathrm{Factor}[\mathrm{eq}3 ~/.~ \{ [c^{'}]^2 \rightarrow \mathrm{auxc}^{'}, [c^{''}]^2 \rightarrow \mathrm{auxc}^{''} \}], 
\end{eqnarray}
where $\mathrm{auxc}^{'}$ and $\mathrm{auxc}^{''}$ are given by Eq.(\ref{auxc}) with weights
labeled by $'$ and $''$, respectively.  

$ \bullet~ \mathrm{Step~ Two} $

We next use the same method to eliminate other underisable powers 
now with the help of the divisor (\ref{INV4}). For example, we can use this divisor to
eliminate the terms that contain powers higher or equal to four
in the weight $\bar{b}$. Denoting the quartic power on $\bar{b}$  
by $\mathrm{aux{\bar{b}}}$ we find that its expression from surface (\ref{surfS2}) is,
\begin{eqnarray}
\label{auxbbar}
\mathrm{aux{\bar{b}}} &=& 
\varepsilon_2^2 b^2[\varepsilon_2 a \bar{b} +a^2 - \varepsilon_1a\bar{b} -\bar{b}^2] 
+ \varepsilon_1\varepsilon_2 b^2[b^2 + \bar{b}(\varepsilon_1 a + \bar{b})] 
-[b^2 + \varepsilon_1 a\bar{b}][b^2 + \bar{b}(\varepsilon_1 a + 2\bar{b})] \nonumber \\
\end{eqnarray}

Considering that highest power in the resulting polynomial 
$\mathrm{eq}4$ on the weight $\bar{b}$ is seven the underisable terms
can be replaced as follows,

\begin{eqnarray}
\mathrm{eq}5 &=& \mathrm{Factor}[\mathrm{eq}4 ~/.~ \{ [{\bar{b}}^{'}]^7 \rightarrow \mathrm{aux{\bar{b}}}^{'}~ [{\bar{b}}^{'}]^3, [{\bar{b}}^{''}]^7 \rightarrow \mathrm{aux{\bar{b}}}^{''}]~ [{\bar{b}}^{''}]^3 \}] \\
\mathrm{eq}6 &=& \mathrm{Factor}[\mathrm{eq}5 ~/.~ \{ [{\bar{b}}^{'}]^6 \rightarrow \mathrm{aux{\bar{b}}}^{'}~ [{\bar{b}}^{'}]^2, [{\bar{b}}^{''}]^6 \rightarrow  \mathrm{aux{\bar{b}}}^{''}~ [{\bar{b}}^{''}]^2 \}] \\
\mathrm{eq}7 &=& \mathrm{Factor}[\mathrm{eq}6 ~/.~ \{ [{\bar{b}}^{'}]^5 \rightarrow \mathrm{aux{\bar{b}}}^{'}~ [{\bar{b}}^{'}], [{\bar{b}}^{''}]^5 \rightarrow \mathrm{aux{\bar{b}}}^{''}~ [{\bar{b}}^{''}] \}] \\
\mathrm{eq}8 &=& \mathrm{Factor}[\mathrm{eq}7 ~/.~ \{ [{\bar{b}}^{'}]^4 \rightarrow \mathrm{aux{\bar{b}}}^{'}, [{\bar{b}}^{''}]^4 \rightarrow \mathrm{aux{\bar{b}}}^{''} \}], 
\end{eqnarray}
where $\mathrm{aux{\bar{b}}}^{'}$ 
and $\mathrm{aux{\bar{b}}}^{''}$ are obtained from (\ref{auxbbar}) 
by using the respective label on the weights. 

$ \bullet~ \mathrm{Step~ Three} $

It turns out that the final polynomial relation $\mathrm{eq}8$  is either automatically zero
or becomes proportional to the factor $\Lambda_0^2-\Lambda_0+1$. In the latter case we can use
the simple substitution, 
\begin{eqnarray}
\mathrm{eqend} = \mathrm{Factor}[\mathrm{eq}8 ~/.~ \Lambda_0^2 \rightarrow \Lambda_0-1] 
\end{eqnarray}

We finally remark that similar verification for the special branch is much simpler 
since the weight $\bar{b}$ can be easily extracted from the divisor (\ref{DIV3}).
In this case the functional relations become dependent only on the weights $a$,
$b$ and $c$. We now use the algebraic plane curve (\ref{curva6}) to extract the
power $c^4$ and denoting it by $\mathrm{buxc}$ we obtain,
\begin{eqnarray}
\label{buxc}
\mathrm{buxc} &=& \frac{a^6 + a^4b^2 + \Lambda_0a^4b^2 + a^2b^4 + 
         \Lambda_0a^2b^4 + \Lambda_0b^6 + \Lambda_4a^3bc^2 + 
         \Lambda_0\Lambda_4ab^3c^2}{a^2 - b^2}
\end{eqnarray}

We find that the highest power on the weight $c$ is always governed by $c^4$ and
thus a given polynomial $\mathrm{eq}[*]$ can be verified
through the steps,
\begin{eqnarray}
\mathrm{eq}1 &=& \mathrm{Factor}[\mathrm{eq}[*]] \\
\mathrm{eq}2 &=& \mathrm{Factor}[\mathrm{eq}1 ~/.~ \{ [c^{'}]^4 \rightarrow ~\mathrm{buxc}^{'}, 
[c^{''}]^4 \rightarrow \mathrm{buxc}^{''} \}] \\
\mathrm{eqend} &=& \mathrm{Factor}[\mathrm{eq}2 ~/.~ \Lambda_0^2 \rightarrow \Lambda_0-1] 
\end{eqnarray}
where $\mathrm{buxc}^{'}$ and $\mathrm{buxc}^{''}$ are determined in terms of Eq.(\ref{buxc}).

\addcontentsline{toc}{section}{Appendix B}
\section*{\bf Appendix B: Elliptic Curves}
\setcounter{equation}{0}
\renewcommand{\theequation}{B.\arabic{equation}}

In this appendix we shall show that the cone (\ref{surfS2}) is
in fact defined over an elliptic curve. To this end it is sufficient
to work in a given affine chart and here we choose the one defined
by setting $\bar{b}=1$. In this affine chart we can 
re-scale the coordinates as follows, 
\EQ
a=x\bar{b}~~\mathrm{and}~~
b=y\bar{b},
\EN
and by substituting this re-scaling of coordinates in
Eq.(\ref{surfS2}) we find that the polynomial 
$S_2(x\bar{b},y\bar{b},\bar{b}) \subset \mathbb{C}[x,y]$ becomes,
\begin{eqnarray}
\label{surfS2A}
S_2(x,y) &=&(\varepsilon_1^2 - \varepsilon_2^2y^2)x^2 +\left[2\varepsilon_1 - (\varepsilon_1^2\varepsilon_2 + \varepsilon_2^3 - 2\varepsilon_1 -\varepsilon_1\varepsilon_2^2)y^2\right]x \nonumber \\
&+& 1+(2 - \varepsilon_1\varepsilon_2 + \varepsilon_2^2)y^2+(1-\varepsilon_1\varepsilon_2)y^4.
\end{eqnarray}

We now can complete the square on the variable $x$ making it possible 
the elimination of the linear term on $x$ of the polynomial (\ref{surfS2A}). This is done at the expense
of adding an extra factor that depends only on the variable $y$ and together with the last term
of Eq.(\ref{surfS2A}) results in a polynomial that factorizes into two pieces. As a result we are able to
define the following one-to-one transformation,
\EQ
\label{JON1}
\begin{array}{ccc}
x & \longmapsto & \frac{R_1(y_1)x_1-R_2(y_1)}{R_3(y_1)}, \\ 
y & \longmapsto & y_1
\end{array}
\EN
where the expressions of the polynomials $R_1(y_1)$, $R_2(y_1)$ and $R_3(y_1)$ are,
\begin{eqnarray}
R_1(y_1) &=& 4\mathrm{i}\varepsilon_2\sqrt{1-\varepsilon_1\varepsilon_2}(\varepsilon_1^2-\varepsilon_2^2y_1^2)y_1,  \nonumber \\
\label{JON2}
R_2(y_1) &=& 2\mathrm{i}(\varepsilon_1^2-\varepsilon_2^2y_1^2)\left[2\varepsilon_1-(\varepsilon_1^2\varepsilon_2+\varepsilon_2^3-2\varepsilon_1-\varepsilon_1\varepsilon_2^2)y_1^2\right], \\
R_3(y_1) &=& 4\mathrm{i}(\varepsilon_1^2-\varepsilon_2^2y_1^2)^2. \nonumber 
\end{eqnarray}

The map (\ref{JON1},\ref{JON2}) is known in the literature as de 
Jonqui\`eres transformation and the zero set of $S_2(x,y)$ 
in the new variables $x_1$ and $y_1$ turns out to be 
equivalent to the curve,
\EQ
\label{curveJA}
x_1^2=y_1^4+\frac{\left[8 + \varepsilon_1^4 - 2\varepsilon_1^3\varepsilon_2 + 4\varepsilon_2^2 + \varepsilon_2^4 - 2\varepsilon_1\varepsilon_2(4 + \varepsilon_2^2) + \varepsilon_1^2(4 + 3\varepsilon_2^2)\right]}{4(1-\varepsilon_1\varepsilon_2)} y_1^2 +1.
\EN

The above polynomial has already the form of an elliptic curve 
since the right-hand side of Eq.(\ref{curveJA})
is a degree four polynomial in $\mathbb{C}[y_1]$. In fact, this curve can easily be brought into
the form of a Jacobi quartic. Let $\pm r_1$ and $\pm r_2$ be the roots of the biquadratic polynomial
on the variable $y_1$. Then by means of the straightforward re-scaling of coordinates,  
$x_1=r_1 r_2 x_2~~\mathrm{and}~~y_1=r_1 y_2$ the plane curve (\ref{curveJA}) can be rewritten as
\EQ
x_2^2=(1-y_2^2)(1-k^2y_2^2),
\EN
whose corresponding modulus parameter is $k=\frac{r_1}{r_2}$. 

We remark that all the above reasoning is valid as long as the discriminant of
corresponding biquadratic polynomial on the variable $y_1$ is not zero.
By direct inspection of the right-hand of Eq.(\ref{curveJA}) one finds that 
the expression of such discriminant is,
\EQ
\label{discri}
\Delta=\left[8 + \varepsilon_1^4 - 2\varepsilon_1^3\varepsilon_2 + 4\varepsilon_2^2 + \varepsilon_2^4 - 2\varepsilon_1\varepsilon_2(4 + \varepsilon_2^2) + \varepsilon_1^2(4 + 3\varepsilon_2^2)\right]^2-64(1-\varepsilon_1\varepsilon_2)^2,
\EN

It turns out that when $\Delta=0$ the elliptic plane 
curve (\ref{curveJA}) can be factorized 
in terms of two conics and therefore the 
original surface (\ref{surfS2}) becomes rational ruled. From Eq.(\ref{discri})
it is not difficult to find this generation occurs 
in the following one-dimensional
submanifolds, 
\begin{eqnarray}
\label{mani1A}
&& 4 - 2\varepsilon_1 + \varepsilon_1^2 - 2\varepsilon_2 - \varepsilon_1\varepsilon_2 + \varepsilon_2^2=0, \\ 
\label{mani1B}
&& 4 + 2\varepsilon_1 + \varepsilon_1^2 + 2\varepsilon_2 - \varepsilon_1\varepsilon_2 + \varepsilon_2^2 =0, \\
&& 4\varepsilon_1^2 + \varepsilon_1^4 - 2\varepsilon_1^3\varepsilon_2 + 4\varepsilon_2^2 + 3\varepsilon_1^2\varepsilon_2^2 - 
       2\varepsilon_1\varepsilon_2^3 + \varepsilon_2^4 =0.
\end{eqnarray}

Note that the above constraints are symmetrical under the exchange of parameters 
$\varepsilon_1 \leftrightarrow \varepsilon_2$ emphasizing that our initial choice of free 
parameters was indeed appropriate. In addition, we observe that 
the first two submanifolds (\ref{mani1A},\ref{mani1B}) 
can be further reduced as the product of linear terms given 
by Eq.(\ref{mani1},\ref{mani2}), respectively.

\addcontentsline{toc}{section}{Appendix C}
\section*{\bf Appendix C: Singularities of Curves}
\setcounter{equation}{0}
\renewcommand{\theequation}{C.\arabic{equation}}

The purpose of this section is to present the technical details concerning 
the singular locus of the degree eight 
algebraic curves discussed in subsection \ref{maincurves}:

$\bullet$ The Curve $\mathrm{C}_1(a,b,c)$

We start by recalling that the singularities of this curve are sited
in the following points,
\EQ
[a_s:b_s:1],~~
[\pm 
\exp(\mathrm{i} \frac{\pi}{3}):1:0],~~ 
[\pm \exp(\mathrm{i} \frac{2\pi}{3}):1:0], 
\EN
where the coordinates $a_s$ and $b_s$ are a subset of solutions of the
relations,
\begin{eqnarray}
\label{singaf1A}
&& \varepsilon_2^2 a_s^3 + (\varepsilon_1 \varepsilon_2-2)a_sb_s^2 + \varepsilon_2 b_s=0, \\
\label{singaf2A}
&& \varepsilon_2^2 b_s^3 +
(2 - \varepsilon_1 \varepsilon_2 + \varepsilon_2^2)a_s^2b_s  + \varepsilon_2 a_s=0.
\end{eqnarray}

The above equations can be solved by first considering the resultant of
the polynomials with respect either to $a_s$ or $b_s$. The resultant has the
merit to eliminate one of the variables and as result we have an univariate
polynomial. Considering the resultant with respect to the coordinate $a_s$
we find,
\EQ
\label{polbs}
b_s\left \{\varepsilon_2^4+\left[\alpha_2-\varepsilon_2^2(4\alpha_1+\varepsilon_2^4)\right]b_s^4 
-\alpha_1^2b_s^8 \right \}=0,
\EN
where the dependence of the coefficients $\alpha_1$ and $\alpha_2$ on the parameters
$\varepsilon_1$ and $\varepsilon_2$ are, 
\begin{eqnarray}
\alpha_1 &=&4 + \varepsilon_2\left[2\varepsilon_2 + \varepsilon_1^2\varepsilon_2 + \varepsilon_2^3 - 
            \varepsilon_1(4 + \varepsilon_2^2)\right], \\
\alpha_2 &=&-16 + 24\varepsilon_1\varepsilon_2 - 4(1 + 3\varepsilon_1^2)\varepsilon_2^2 + 
     2\varepsilon_1(2 + \varepsilon_1^2)\varepsilon_2^3 - (4 + \varepsilon_1^2)\varepsilon_2^4 
+ 2\varepsilon_1\varepsilon_2^5 + \varepsilon_2^6. 
\end{eqnarray}

The trivial solution $b_s=0$ to Eq.(\ref{polbs}) implies also $a_s=0$ which has
to be discarded since the point $[0:0:1]$ does not belong to the curve $\mathrm{C}_1(a,b,1)$.
This means that the allowed values for the coordinate $b_s$ are the eight roots of the
second factor of the polynomial (\ref{polbs}). The corresponding coordinates for $a_s$
can now be obtained by applying similar reasoning we used to intersect the surfaces
in subsection \ref{maincurves}. They can be expressed in terms of the variables
$b_s$ by the following expression,  
\EQ
a_s= \left[\frac{b_s}{\varepsilon_2}\right]^3 
\frac{\left\{\alpha_3+[\varepsilon_2(\varepsilon_1-\varepsilon_2)-2]\alpha_1^2b_s^4\right\}}{\alpha_1+\varepsilon_2^2[2-\varepsilon_2(\varepsilon_1-\varepsilon_2)]},
\EN
where the coefficient 
$\alpha_3$ is given by, 
\begin{eqnarray}
\alpha_3 &=& -32 + 64\varepsilon_1\varepsilon_2 - 8(7 + 6\varepsilon_1^2)\varepsilon_2^2 + 
     4\varepsilon_1(21 + 4\varepsilon_1^2)\varepsilon_2^3 - 2(24 + 21\varepsilon_1^2 + \varepsilon_1^4)\varepsilon_2^4  \nonumber \\
     &+& \varepsilon_1(48 + 7\varepsilon_1^2)\varepsilon_2^5 
-2(11 + 6\varepsilon_1^2)\varepsilon_2^6 + 
     11\varepsilon_1\varepsilon_2^7 - 5\varepsilon_2^8.
\end{eqnarray}

$\bullet$ The Curve $\mathrm{Q}_1(x,y,z)$

By solving the polynomial equations associated to the singular 
locus (\ref{sing}) of the target curve of the double
cover map $\phi$ we find nine singular points. Five of them are 
located on the affine plane and they are given by
\EQ
\mathrm{P}_{A}=[0:0:1]~\mathrm{and}~[x_s:y_s:1],
\EN
where the coordinates $x_s$ and $y_s$ satisfy the following decoupled equations, 
\begin{eqnarray}
&& \alpha_1^2x_s^4+\alpha_2x_s^2 -\varepsilon_2^4=0, \\
&& y_s=\frac{\varepsilon_2(2-\varepsilon_1\varepsilon_2) -\varepsilon_2\alpha_1x_s^2}
{4-4\varepsilon_1\varepsilon_2+\varepsilon_1^2\varepsilon_2^2+\varepsilon_2^4}.
\end{eqnarray}

The singularities at the infinity line $z=0$ sit on the 
same places of corresponding
singular points associated to 
the domain degree eight curve, that is 
\EQ
\mathrm{P}_{\infty}=[\pm 
\exp(\mathrm{i} \frac{\pi}{3}):1:0],~~ 
[\pm \exp(\mathrm{i} \frac{2\pi}{3}):1:0], 
\EN
except that now they behave as ordinary double points.

The only singularity that is not an ordinary double point turns out to be the one
sited at the origin of the affine plane
$P_0=[0:0:1]$. The respective index of multiplicity is $m_{P_0}=4$ and it has an extra
neighboring infinitesimal singularity. The desingularization diagram is
thus given by, 
\EQ
\label{blow1}
\renewcommand{\arraystretch}{1.5}
\begin{array}{ccccc}
{{\bf{{Q}}}}_1 & \overset{\bar{\pi}_2}{\longrightarrow} & \tilde{\mathrm{Q}}_{1} 
& \overset{\bar{\pi}_1}{\longrightarrow} & \mathrm{Q}_1(a,b,c)  
\end{array}
\EN
where the curve $\tilde{\mathrm{Q}}_1$ carries the infinitely near singularity
associated to the point $P_0=[0:0:1]$  whose index of multiplicity is also four.
By using this information we can easily obtain the corresponding genus of
the normalization ${\bf Q}_1$, see Eq.(\ref{genus1}).

\addcontentsline{toc}{section}{Appendix D}
\section*{\bf Appendix D: Reducible Curves}
\setcounter{equation}{0}
\renewcommand{\theequation}{D.\arabic{equation}}

The purpose of this Appendix is to present the explicit expressions of the plane curves resulting
from the intersection of the surfaces (\ref{surfS1},\ref{surfS2}) when the parameters
$\varepsilon_1$ and $\varepsilon_2$ are restricted to the 
submanifolds (\ref{mani1}-\ref{mani3}). We shall also see such 
degeneration gives origin to singular
algebraic quartic curves with genus one.

$\bullet$~The linear submanifolds:

We first notice that it is enough to consider one of the 
submanifolds (\ref{mani1},\ref{mani2})
because they are trivially related under the transformation
$\varepsilon_1 \rightarrow -\varepsilon_1$ and 
$\varepsilon_2 \rightarrow -\varepsilon_2$. In the case of the submanifold (\ref{mani1}) we find,
after eliminating the parameter $\varepsilon_2$, that the octic plane curve (\ref{curva8})
becomes factorizable in terms of the following product, 
\EQ
\mathrm{C}_1(a,b,c) =g(a,b,c) g(-a,b,\mathrm{i}c)
\EN
where the expression of the degree four plane curve $g(a,b,c)$ is,
\begin{eqnarray}
\label{quartic1}
g(a,b,c)&=& (-\frac{1}{2} \pm \mathrm{i}\frac{\sqrt{3}}{2} + \varepsilon_1)a^4  \pm
     (-1 \mp \mathrm{i}\sqrt{3} - \varepsilon_1 \pm \mathrm{i}\sqrt{3}\varepsilon_1)a^3b + 
     (\frac{3}{2} \pm \mathrm{i}\frac{\sqrt{3}}{2} \mp \mathrm{i}\sqrt{3}\varepsilon_1)a^2b^2 \nonumber \\ &\pm& 
     (-2 + \varepsilon_1 \pm \mathrm{i}\sqrt{3}\varepsilon_1)ab^3 + (\frac{1}{2} \mp \mathrm{i}\frac{\sqrt{3}}{2} - \varepsilon_1)b^4  \pm
     (-1 \pm \mathrm{i}\sqrt{3} + \varepsilon_1)a^2c^2 \nonumber \\
&+& (-2 + \frac{\varepsilon_1}{2} \pm \mathrm{i}\frac{\sqrt{3}}{2}\varepsilon_1)ab
      c^2 \pm (-1 \mp \mathrm{i}\sqrt{3} - \frac{\varepsilon_1}{2} \pm \mathrm{i}\frac{\sqrt{3}}{2}\varepsilon_1)b^2c^2 + 
     (-\frac{1}{2} \pm \mathrm{i}\frac{\sqrt{3}}{2})c^4
\end{eqnarray}
such that the symbol $\pm$ refer to the two possible signs of the submanifold (\ref{mani1}).

The plane curve (\ref{quartic1}) has only one singular point sited at the infinity line $c=0$ 
which behaves as a tacnode 
and as a consequence of that it has genus one. This means that there exists room for emerging
new singularities for particular values of the parameter $\varepsilon_1$
and an extra generation to a rational plane curve may be possible. Indeed, we find that this happens
at the following values of the quadratic submanifolds (\ref{mani1},\ref{mani2}),
\EQ
\label{specialp}
\varepsilon_1=\varepsilon_2=\pm 2,~~~\mathrm{and}~~~
\varepsilon_1=-\varepsilon_2=\pm \frac{2 \mathrm{i}}{\sqrt{3}},
\EN
where now the quartic plane curves factorizes once again in the product of two conics. As a result, the original
octic plane curve (\ref{curva8}) degenerates to genus zero curve and the corresponding weights of the
vertex model becomes trigonometric 
at the specific points (\ref{specialp}).

$\bullet$~The quartic submanifold:

The essential ingredient in the analysis of the third submanifold (\ref{mani3}) 
is to note that we are in fact dealing with a curve that can be rationally parameterized.
In fact, one  easily finds that submanifold (\ref{mani3}) has
the maximal number of three ordinary singular points when 
viewed in the projective space. Now given such three double points on a quartic curve
we can fix any other point on it and pass conics through these four points. By using
this family of conics it is well known that one can establish a rational parameterization 
and the final result is,
\begin{eqnarray}
\label{param}
\varepsilon_1&=& -\frac{4(-3+14\kappa)(147t^2-98t+294\kappa t
+12-46\kappa)(-49t^2+98\kappa t
+2-24\kappa)}{(133t^2-56t+84\kappa t+4-6
\kappa)(343t^2-196t+980\kappa t+24-190
\kappa)} \\ \nonumber \\
\varepsilon_2&=& -\frac{2(-4+63\kappa)(147t^2-98t+294\kappa t
+12-46\kappa)(637t^2-294t+490\kappa t
+36-138\kappa)}{13(133t^2-56t+84\kappa t+
4-6\kappa)(343t^2-196t+980\kappa t
+24-190\kappa)} \nonumber \\
\end{eqnarray}
where $t$ is the parameterization variable and $\kappa$ is a constant factor
$\kappa=\frac{6\pm\mathrm{i}\sqrt{13}}{49}$.

With the help of the parameterization (\ref{param}) we then are able to investigate the
explicit factorization of the octic plane curve (\ref{curva8}) in terms of the product of
two quartic curves, namely 
\EQ
\mathrm{C}_1(a,b,c) =h(a,b,c) h(-a,b,\mathrm{i}c)
\EN
where the expression for $h(a,b,c)$ is,
\begin{eqnarray}
\label{quartic2}
h(a,b,c)& =& \frac{\kappa_1}{109531219}(a^4+b^4+a^2b^2)+ \frac{(29\pm 36\mathrm{i}\sqrt{13})\kappa_2}{14567652127}c^4
\mp\frac{2(\pm 2 \mathrm{i} \sqrt{13}+9)\kappa_3^2}{6900466797}a^2c^2 \nonumber \\
&\pm& \frac{2(\pm 2 \mathrm{i} \sqrt{13}+9)\kappa_4^2}{1684480617001}b^2c^2
\mp\frac{2(\pm 9 \mathrm{i} \sqrt{13}-26)\kappa_3 \kappa_4}{388726296231}abc^2
\end{eqnarray}
such that the coefficients $\kappa_1, \cdots, \kappa_4$ are given by,
\begin{eqnarray}
\kappa_1 &=& (45619t^2- 16856t \pm 392 \mathrm{i}  \sqrt{13}t+ 1836 \mp 86 \mathrm{i} \sqrt{13})(160 - 1372t + 2401t^2 \mp 6 \mathrm{i} \sqrt{13})\nonumber  \\
\kappa_2 &=& (16807t^2 \pm 980 \mathrm{i} \sqrt{13}t - 3724 t \mp 190 \mathrm{i} \sqrt{13}+ 36)(160 - 2240t \pm 84 \mathrm{i} \sqrt{13}t+ 6517t^2 \mp 6\mathrm{i} \sqrt{13}) \nonumber \\
\kappa_3 &=& (147 t \pm 3\mathrm{i} \sqrt{13}-31 \mp 12\mathrm{i} + 2\sqrt{13})(147t-31 \pm 12\mathrm{i} - 2\sqrt{13} \pm 3 \mathrm{i} \sqrt{13}) \nonumber \\
\kappa_4 &=& (637t \pm 5\mathrm{i} \sqrt{13}- 117 \mp 26\mathrm{i} - 12\sqrt{13})(637t \pm 5 \mathrm{i} \sqrt{13}- 117 \pm 26 \mathrm{i} + 12\sqrt{13})
\end{eqnarray}

For generic values of the free variable $t$ one finds that 
the quartic plane curve (\ref{quartic2}) has
two ordinary singular points and thus has again genus one. We remark that in this submanifold we have not been able to find
a further generation to rational curves. This however can not be ruled out since in this case the analysis is more subtle.

\end{document}